\documentclass[prd,aps,eqsecnum,floatfix,nofootinbib,twocolumn,tightenlines,preprintnumbers,superscriptaddress]{revtex4}
\usepackage{dcolumn}% Align table columns on decimal point
\usepackage{amsmath}
\usepackage{latexsym}
\usepackage{graphicx}
\usepackage{bm}
\usepackage{colordvi}
\usepackage{xcolor}
\def\a{\alpha}
\def\b{\beta}

\def\d{\delta}
\def\e{\epsilon}                % Also, \varepsilon
\def\g{\gamma}

\def\m{\mu}

\def\s{\sigma}                  %       \varsigma
\def\t{\tau}

\def\co{{\cal O}}

\def\bx{{\bf x}}

\def\svev#1{\left\langle #1\right\rangle}       % variable < >
\def\sket#1{\left| #1\right\rangle}             % variable | >
\def\Tr{{\rm Tr}\,}

\long \def \blockcomment #1\endcomment{}
\def\det{{\rm det}}

\def\Eq#1{Eq.~(\ref{#1})}

\def\meff{m_{\text{eff}}}

\def\xhat{\hat{x}}
\def\ubar{\bar{u}}
\def\dbar{\bar{d}}
\def\SU{\text{SU}}
\def\SO{\text{SO}}

\newcommand{\bee}{\begin{equation}}
\newcommand{\ee}{\end{equation}}
\newcommand{\beea}{\begin{eqnarray}}
\newcommand{\eea}{\end{eqnarray}}

\begin{document}
%%%%%%%%%%%%%%%%%%%%%%%%%%%%%%%%%%%%%%%%%%%%%%%%%%%%%%%%%%%%%%%%%%%%%%
\title{
Spectroscopy of SU(4) gauge theory with two flavors of sextet fermions}
\author{Thomas DeGrand}
\author{Yuzhi Liu}
\affiliation{Department of Physics,
University of Colorado, Boulder, CO 80309, USA}
\author{Ethan T.~Neil}
\affiliation{Department of Physics,
University of Colorado, Boulder, CO 80309, USA}
\affiliation{RIKEN-BNL Research Center, Brookhaven National Laboratory,
Upton, NY 11973, USA}
\author{Yigal Shamir}
\affiliation{Raymond and Beverly Sackler School of Physics and Astronomy,
Tel~Aviv University, 69978 Tel~Aviv, Israel}
\author{Benjamin Svetitsky}
\affiliation{Raymond and Beverly Sackler School of Physics and Astronomy,
Tel~Aviv University, 69978 Tel~Aviv, Israel}
\affiliation{Yukawa Institute for Theoretical Physics, Kyoto University, Kyoto 606-8502, Japan}

\begin{abstract}
We present a first look at the spectroscopy of SU(4) gauge theory coupled to two flavors
of Dirac fermions in the two-index antisymmetric representation, which is a real representation.
We compute meson and diquark masses, the pseudoscalar and vector meson
decay constants, and the masses of six-quark baryons.
We make comparisons with large-$N_c$ expectations.
\end{abstract}

\preprint{COLO-HEP-586}
\preprint{YITP-15-4}
%\pacs{11.15.Ha}
%\keywords{Suggested keywords}
\maketitle

%%%%%%%%%%%%%%%%%%%%%%%%%%%%%%%%%%%%%%%%%%%%%%%%%%%%%%%%%%%%%%%%%%%%%
\section{Introduction}
%%%%%%%%%%%%%%%%%%%%%%%%%%%%%%%%%%%%%%%%%%%%%%%%%%%%%%%%%%%%%%%%%%%%%
Composite Higgs models \cite{Perelstein:2005ka, Giudice:2007fh, Barbieri:2007bh, Contino:2010rs, Bellazzini:2014yua} are frequently based on nonlinear sigma models.
The most straightforward ultraviolet completion of such a model is a gauge theory with the corresponding spontaneous breaking of global symmetries.
A symmetry-breaking scheme that is much discussed is $\text{SU}(N)\to\text{SO}(N)$\@.
Such a breaking scheme can be accommodated in an SU($N_c$) gauge theory where the fermions are in a real representation of the gauge group.
Then $N=2N_f$, where $N_f$ is the number of flavors of Dirac fermions.

As the first stage in a program of investigating gauge theories of interest beyond the Standard Model, we focus here on the SU(4) gauge theory with fermions in the two-index antisymmetric representation (denoted AS2 henceforth).
This is the sextet of SU(4), a real representation.
We choose $N_f=2$ flavors of Dirac fermions, so that the global chiral symmetry is also SU(4), which we expect to see spontaneously broken to SO(4).

This theory is a way station on the route to the SU(4) gauge theory with five Majorana fermions.
That theory is the most economical way to realize the symmetry breaking of an SU(5)/SO(5) sigma model, which is the basis of, for instance, the littlest Higgs model~\cite{ArkaniHamed:2002qy}.
The $\SU(5)/\SO(5)$ sigma model is also central to more recent
composite-Higgs models~\cite{Vecchi:2013bja,Ferretti:2014qta,Ferretti:2013kya}.
Indeed, Vecchi~\cite{Ferretti:2013kya} argued that the
SU(4) theory with AS2 fermions is the most attractive candidate
within this approach;  Ferretti~\cite{Ferretti:2014qta}
elaborated on the phenomenology of this composite-Higgs model.%
\footnote{The models of Refs.~\cite{Ferretti:2014qta,Ferretti:2013kya}
include fermions in the fundamental representation in addition
to the AS2, in order to give the top quark a mass via
the partial-compositeness mechanism \cite{Kaplan:1991dc}.  See also Ref.~\cite{Golterman:2015zwa}.}  
We can simulate the $N_f=2$ theory with the standard 
Hybrid Monte Carlo (HMC) algorithm, while study of the theory
with 5 Majorana fermions will require the more expensive rational HMC algorithm.
 
In this paper we present a study of the basic features of the $N_f=2$ theory, namely, its phase
 diagram and spectrum; preliminary results were presented in Ref.~\cite{DeGrand:2014cea}.
The spectrum must exhibit multiplets of the unbroken SO(4) flavor symmetry.
One feature of these multiplets is that mesons and diquarks transform into each other under SO(4).
Because of this, baryons with more than two quarks are of particular interest;
for reasons to be stated below, we study baryons made of six quarks.  

We could simply present our results for spectroscopy without further analysis. However, we feel that,
rather than just doing that, we should try to give them some context: we are studying
a confining, chirally broken system. How are the masses and matrix elements we compute
for our system different from, or similar to, what is seen in other confining and chirally broken systems?

The context we choose to use is the $1/N_c$ expansion.
Theories with fermions in two-index representations have been studied extensively in a 
$1/N_c$ framework~\cite{Corrigan:1979xf}, as an alternative to the original $1/N_c$ 
expansion that deals with fermions in the fundamental representation~\cite{'tHooft:1973jz,'tHooft:1974hx}. 
AS2 fermion loops are not suppressed at large $N_c$, leading to different systematics 
than the conventional $1/N_c$ expansion.

Either $1/N_c$ expansion can in principle be applied to QCD, since for $N_c=3$ the AS2 and 
fundamental representations are isomorphic.  Furthermore, interesting equivalences to 
supersymmetric Yang--Mills theory in the $N_c\to\infty$ limit have been argued for theories 
with AS2 fermions \cite{Armoni:2003fb,Armoni:2003gp,Armoni:2004uu, Armoni:2009zq}, related to
 the orientifold equivalence among all gauge theories with two-index representations
 (adjoint, AS2, or symmetric) in the large-$N_c$ limit \cite{Kovtun:2003hr,Unsal:2007fb,Hanada:2011ju}.
This framework continues to attract interest.\footnote{
See for example Refs.~\cite{Bolognesi:2006ws,Cherman:2006iy,Cherman:2009fh,Cohen:2009wm,Cherman:2012eg,
 Cohen:2014via,Buisseret:2011aa}.
A recent review with emphasis on properties of baryons is given in Ref.~\cite{Matagne:2014lla}.}

Our new data on the spectrum and decay constants for SU(3) and SU(4) theories with AS2 fermions will 
allow us to make a qualitative comparison to the scaling predictions of this alternative $1/N_c$ expansion. 
 This comparison is made at a single value of the bare gauge coupling, in the confined and chirally broken
 phase of our lattice action.
We have in hand already-published spectroscopy data for the SU(3), SU(5), and SU(7) gauge theories with
 fermions in the fundamental representation.
These previous studies were performed in the quenched approximation, but a comparison to our new
 dynamical-fermion data for the SU(3) theory, roughly matched in simulation volume and lattice 
spacing to the quenched data sets, shows little effect of quenching.
This is because the dynamical fermion masses are not light enough to produce appreciable differences 
through loop effects.
Of course, for the study of the large-$N_c$ limit with AS2 fermions, quenching is completely unjustified.

Baryons in large $N_c$ are of longstanding interest in the traditional framework with 
fundamental-representation fermions.
They can be analyzed as many-quark states \cite{Witten:1979kh} or can be taken to be topological 
objects in effective theories of
mesons~\cite{Witten:1983tx,Adkins:1983ya,Gervais:1983wq,Gervais:1984rc}.
Large-$N_c$ mass formulas for baryons have been presented in
Refs.~\cite{Jenkins:1993zu,Dashen:1993jt,Dashen:1994qi,Jenkins:1995td,Dai:1995zg,Cherman:2012eg}.
An old review~\cite{Manohar:1998xv} summarizes much of this classic work.

Several recent lattice studies of baryon spectroscopy have touched on large-$N_c$ considerations.
The first \cite{Jenkins:2009wv} was a comparison of lattice Monte Carlo data for $N_c=3$ baryons to large-$N_c$  formulas.
There are also three related studies at $N_c=3$, 5, and~7, comparing quenched spectroscopy with $N_f=2$ flavors of degenerate valence quarks \cite{DeGrand:2012hd}; 
spectroscopy with $N_f=3$ flavors (two degenerate ones and a heavier strange quark) \cite{DeGrand:2013nna};
and $N_f=2$ data to baryon chiral perturbation theory \cite{Cordon:2014sda}.
Finally, Ref.~\cite{Appelquist:2014jch} reports calculations of quenched baryon spectroscopy in SU(4).
The results of these studies all conform to large-$N_c$ expectations for the fundamental representation;
 we will make comparisons of our spectrum to these results where appropriate.
 
 We note in passing that the behavior of AS2 theories at finite baryon density has also attracted some theoretical interest \cite{Frandsen:2005mb, Buchoff:2009za}.
The SU(4) gauge theory is particularly useful for lattice work: Since the AS2 representation is real, the theory at finite baryon density presents no sign problem.

Now we proceed to the body of the paper.
Section \ref{sec:as2} collects some useful group-theoretic results about AS2 fermions in the SU(4) gauge theory, and their special symmetries.
Six-quark baryons emerge as objects of interest.
We present the lattice action and a new discretization issue in Sec.~\ref{sec:action}.
The choice of parameters used for spectroscopy was made after
a scan of the bare parameter space (bare gauge coupling and hopping parameter).
This scan revealed some of the phase structure of this system, which we present in Sec.~\ref{sec:phase}.
We describe our methods for obtaining spectra in Sec.~\ref{sec:spectra} and display 
tables of the resulting meson and baryon spectra.
We then plot these results and offer comparisons among the SU(4) AS2 theory; the SU(3) 
theory; and quenched SU(3), SU(5), and SU(7) theories: for mesons in  Sec.~\ref{sec:mesons} and for baryons in  Sec.~\ref{sec:baryons}.
Finally, Sec.~\ref{sec:conclusions} makes some phenomenological observations, summarizes our results, and suggests future directions.

%%%%%%%%%%%%%%%%%%%%%%%%%%%%%%%%%%%%%%%%%%%%%%%%%%%%%%%%%%%%%%%%%%%%%
\section{Group theory and symmetries \label{sec:as2}}
%%%%%%%%%%%%%%%%%%%%%%%%%%%%%%%%%%%%%%%%%%%%%%%%%%%%%%%%%%%%%%%%%%%%%

In this section we discuss the symmetry aspects of AS2 fermions in SU(4), specializing to $N_f=2$.
In Sec.~\ref{sec:real} we review some basic properties of real and pseudoreal
representations and how they are reflected in symmetries
of the Wilson--Dirac operator and meson/diquark propagators.

In Sec.~\ref{sec:ssb} we turn to global symmetries. The pattern of chiral symmetry
 breaking in SU(4) AS2 is different from that in SU(3)
 gauge theories
with fundamental-representation fermions, because the AS2 fermions
live in a real representation of the gauge group. The usual breaking pattern, 
$\SU(N_f)\times \SU(N_f) \rightarrow
\SU(N_f)$, is replaced by $\SU(2N_f) \rightarrow \SO(2N_f)$~\cite{Peskin:1980gc,Preskill:1980mz,Kosower:1984aw}.
There are $2N_f^2+N_f-1$ Nambu--Goldstone bosons (NGBs), nine in all for $N_f=2$.
A consequence of reality is that, in addition to meson ($q \bar q$) and baryon states, there are
also diquark states.
Symmetries associated with the fermions' reality means that all
 diquark correlators are identical
to corresponding meson ones. For example, the diquarks are needed to fill out the NGB multiplets.
The nine NGBs
consist of three isotriplets: one multiplet is $q \bar q$, one is $q  q$ and one is $\bar q \bar q$.

The global symmetry of the $N_f=2$ AS2 theory is thus $\SU(2N_f)=\SU(4)$.
After dynamical symmetry breaking, the unbroken symmetry is SO(4).
We elaborate on this symmetry-breaking pattern,
focusing on how the two invariant SU(2) subgroups of SO(4) are
realized.  As an example, we classify the nine Nambu--Goldstone bosons
under the unbroken symmetry.  In Sec.~\ref{sec:B} we  recall the equivalence between 
the AS2 representation of color SU(4)
and the vector representation of SO(6).  We use this equivalence to introduce
the color-singlet state of six AS2 fermions, which is fully antisymmetric
in its color indices.
The remainder of the section explains why the six-quark baryons are the interesting 
baryonic states and describes how we construct baryon operators.

%%%%%%%%%%%%%%%%%%%%%%%%%%%%%%%%%%%%%%%%%%%%%%%%%%%%%%%%%%%%%%%%%%%%%%%%%%%%
\subsection{Symmetries of real and pseudoreal representations \label{sec:real}}
%%%%%%%%%%%%%%%%%%%%%%%%%%%%%%%%%%%%%%%%%%%%%%%%%%%%%%%%%%%%%%%%%%%%%%%%%%%%

We begin by recalling how basis states of the AS2 irreducible representation
are built from color basis states $\sket{i}$ in the fundamental representation.
The basis states of the antisymmetric representation are
\bee
  \sket{ij} = \frac{1}{\sqrt{2}}\,
  \Big(\sket{i}\sket{j} - \sket{j}\sket{i}\Big)\,,
  \qquad  1\le i<j\le N_c\,.
\label{eq:basis}
\ee
There are $N_c(N_c-1)/2$ basis states---six states for $N_c=4$.
Starting from the transformation rule
$\sket{i}' = \sum_k U_{ik}\sket{k}$,
the AS2 states transform as
\beea
  \sket{ij}' &=&
  \frac{1}{\sqrt{2}}\,
  \Big(\sket{i}'\sket{j}' - \sket{j}'\sket{i}'\Big)
\nonumber\\
  &=&
  \sum_{k<l} \Big( U_{ik} U_{jl} - U_{jk} U_{il} \Big) \sket{kl}
\nonumber\\
  &\equiv&
  \sum_{k<l} U_{[ij][kl]} \sket{kl}  \,.
\label{eq:recouple}
\eea
Equation (\ref{eq:recouple}) provides the composition rule for $U_{[ij][kl]}$, the link
matrices in the AS2 representation, in terms of links $U_{ik}$
which are elements of the fundamental representation of SU($N_c$).

The AS2 representation of SU(4) is real.
Correlators of fermionic bilinears reflect this reality.
This is similar to the situation in $N_c=2$
where the fundamental representation is pseudoreal.
As in the case of the two-color theory with fundamental fermions,
which was nicely described in Ref.~\cite{Lewis:2011zb},
for any (pseudo)real representation there is an exact identity between
a meson correlator and  a corresponding diquark correlator,
\begin{align}
   &\left\langle \bar u(x)\Gamma         d(x) \,
                \bar d(y)\Gamma^\dagger u(y)\right\rangle \nonumber \\ 
   &\hspace{4mm}=  \left\langle \bar u(x)              \Gamma \left(S C \bar d(x)^T\right)\,
                   \left(d(y)^T S C\right)\Gamma^\dagger        u(y) \right\rangle \ ,
\label{eq:relate}
\end{align}
where $u(x)$ and $d(x)$ are the two flavors of Dirac fermions and $\Gamma$ is a Dirac matrix.
Let us see how this comes about.

The matrix $S$ is defined as follows.
A real or pseudoreal irreducible representation is self-conjugate, meaning that there 
is a quadratic form $S$ such that for two vectors $a$ and $b$  the product $a^TSb$ is a singlet.
Demanding invariance under $a=Ua'$, $b=Ub'$, we find
\bee
U^TSU=S,
\label{eq:Uthru}\ee
which implies that the Hermitian generators $T_a$ satisfy
\bee
T^T_a S = T^*_a S = -S T_a.
\label{eq:Tthru}
\ee
The entries of S are real, $S^* = S$, and it satisfies $S^{-1}=S^T$.
For a real representation $S \equiv R=R^T$ is symmetric, whereas for a pseudoreal
representation  $S \equiv P=-P^T$ is antisymmetric.
For the AS2 representation of SU(4), it is realized as
\bee
S \psi_{[ij]} = \sum_{k<l} \epsilon_{ijkl} \psi_{[kl]}.
\ee
(Note that $\epsilon_{ijkl} = \epsilon_{klij}$, so $S$ is symmetric
as it should be for a real representation.)

The matrix $C$ occurring in \Eq{eq:relate} is the usual
charge-conjugation matrix, which satisfies $C\g_\m = -\g_\m^T C$,
and $C^{-1}=C^\dagger = C^T = -C$.  We recall that charge-conjugation symmetry
acts as\footnote{
  The Euclidean rules (\ref{eq:cconga}) and (\ref{eq:ccongb}) are consistent
  with the Minkowskian relation $\bar \psi = \psi^\dagger \gamma_0$,
  where we have identified $\gamma_0\equiv\gamma_4$.
}
\begin{subequations}
\label{eq:ccong}
\beea
  \psi &\to & C\, \bar \psi^T \ ,
\label{eq:cconga}\\
  \bar \psi  &\to& \psi^T C \ ,
\label{eq:ccongb}\\
  A_\m &\to & -A_\m^*  \hspace{5ex} \mbox{(continuum)} \ ,
\label{eq:ccongc}\\
  U_\m &\to& U_\m^*  \hspace{7ex} \mbox{(lattice)}  \ .
\label{eq:ccongd}
\eea
\end{subequations}

We are now ready to derive the identity~(\ref{eq:relate}).
Consider any fermion action that is invariant under the charge-conjugation
symmetry~(\ref{eq:ccong}) when the fermions belong to a complex representation.
If we now take the fermions to be in a real representation ($S \equiv R$),
then the fermion action will be invariant under the following
discrete symmetry:
\begin{subequations}
\label{eq:ccongR}
\beea
  \psi &\to & R C\, \bar \psi^T \ ,
\label{eq:ccongRa}\\
  \bar \psi &\to& \psi^T C R \ ,
\label{eq:ccongRb}\\
  A_\m &\to & A_\m  \hspace{7ex} \mbox{(continuum)} \ ,
\label{eq:ccongRc}\\
  U_\m &\to& U_\m \hspace{7ex} \mbox{(lattice)}  \ .
\label{eq:ccongRd}
\eea
\end{subequations}
Thanks to the reality condition~(\ref{eq:Uthru}),
the inclusion of $R$ in the fermions' transformation rule
makes up for the fact that the gauge field does not transform.
For a pseudoreal representation ($S \equiv P$), the discrete symmetry is
\begin{subequations}
\label{eq:ccongP}
\beea
  \psi &\to & P C\, \bar \psi^T \ ,
\label{eq:ccongPa}\\
  \bar \psi &\to& -\psi^T C P \ .
\label{eq:ccongPb}
\eea
\end{subequations}
We may apply
the transformation~(\ref{eq:ccongR}) [or~(\ref{eq:ccongP})] to a single Dirac fermion.
This is unlike the usual charge conjugation~(\ref{eq:ccong}),
which acts on the gauge field as well
and must be applied to all fields simultaneously.
For both real and pseudoreal representations, it follows that
the (lattice) Dirac operator satisfies the identity
\bee
  S C D^T  S^{-1} C^{-1} = - SC D^T S^{-1}C = D ,
\label{eq:realDT}
\ee
and \Eq{eq:relate} follows.

We comment in passing
that for Wilson fermions, $\g_5 D^\dagger \g_5 = D$.
Together with (the Hermitian conjugate of) \Eq{eq:realDT} this implies
\bee
  S \gamma_5 C D^* S \gamma_5 C = -D \ ,
\ee
and hence that the fermion determinant is real.

%%%%%%%%%%%%%%%%%%%%%%%%%%%%%%%%%%%%%%%%%%%%%%%%%%%%%%
\subsection{Unbroken SO(4) symmetry and baryon number \label{sec:ssb}}
%%%%%%%%%%%%%%%%%%%%%%%%%%%%%%%%%%%%%%%%%%%%%%%%%%%%%%

In a gauge theory with $N_f$ Dirac fermions in a real representation,
the global symmetry is SU($2N_f$).
After chiral symmetry breaking,
the unbroken symmetry is  SO($2N_f$).
These statements are most obvious when the theory is formulated in terms of
Majorana fermions.  Invariance under the transformation~(\ref{eq:ccongR}) allows
 each Dirac field to be broken up into two Majorana fields, with no mixing in the 
action as long as there are no mass terms.
The number of independent Majorana (or Weyl) fields
is $N_{\text{Maj}}=2N_f$, making the global symmetry SU($2N_f$).
The fermion condensate is a Majorana-fermion bilinear which,
for a real representation, is symmetric in its color indices.
As the expectation value of a scalar operator,
it is antisymmetric in its spin indices, and so it must be
symmetric in its (Majorana) flavor indices.  It then follows
that the unbroken symmetry is SO($2N_f$)
\cite{Peskin:1980gc,Preskill:1980mz,Kosower:1984aw}.

Since we elect to work with two AS2 Dirac fermions (instead of four
Majorana fermions), we should
understand how the SO(4) unbroken symmetry is realized on them.
SO(4) is doubly covered by $\SU(2)\times \SU(2)$.  We will now work
out how the two SU(2) groups act on our Dirac fermions.
As we will see, one of the SU(2) groups may be identified with isospin,
while the baryon number symmetry becomes a subgroup of the other SU(2).

We start with the observation that SO(4) is the symmetry group of the 3-sphere $S^3$,
which in turn can be identified with the SU(2) group manifold via
$\xhat = x_4 + i \sum_{a=1}^3 x_a \s_a$, where $\s_a$ are the Pauli matrices
and $\sum_{\m=1}^4 x_\m^2=1$.  The product group $\SU(2)\times \SU(2)$
is then realized as\footnote{
  The product-group elements $g=-1,$ $h=1$, and $g=1,$ $h=-1$,
  coincide when they act on $\xhat$.  Hence $\SU(2)\times \SU(2)$
  is a double covering of SO(4).
}
\bee
  \xhat \to g \xhat h^\dagger\ , \qquad g,h \in \SU(2) \ .
\label{O4SU22}
\ee

To keep track of the SU(2) transformation properties, 
it is convenient to rearrange the four real coordinates
into two complex ones.  We choose
$\phi_1 = x_4+ix_3$, $\phi_2 = -x_2+ix_1$, so that
\bee
  \xhat = \left(\begin{array}{cr}
    \phi_1 & -\phi_2^* \\
    \phi_2 & \phi_1^*
    \end{array} \right)\ , \qquad
  -\xhat^\dagger = \left(\begin{array}{rc}
    -\phi_1^* & -\phi_2^* \\
    \phi_2 & -\phi_1
    \end{array} \right)\ .
\label{lmult}
\ee
(The minus sign in front of $\xhat^\dagger$ is introduced for convenience below.)

The transformation properties under left-multiplication
are now obvious.  The left column of the $\xhat$ matrix
is an SU(2) doublet $(\phi_1,\phi_2)$.
Denoting this doublet as $\Phi_\a$,
the right column is $\Phi'_\a = \e_{\a\b}\Phi_\b^*$,
which again transforms in the fundamental representation of SU(2).

Next, to obtain the behavior under right-multiplication
we consider the left-action of $h$ on $-\xhat^\dagger$.
We read off the right-multiplication doublets:
$(-\phi_1^*,\phi_2)$ from the left column of $-\xhat^\dagger$,
and $(-\phi_2^*,-\phi_1)$ from its right column.
The left and right doublets are related by interchanging
$\phi_1$ with $-\phi_1^*$.

We now turn to our AS2 theory.  The role of real coordinates is played by
Majorana fermions, whereas that of complex coordinates is played
by Dirac fermions.  What takes the place of complex conjugation is the
transformation~(\ref{eq:ccongR}).
We may arrange our two Dirac fermions, $u$ and $d$, as well as
their antifermions, in complete analogy with Eq.~(\ref{lmult}),
\bee
  \Psi = \left(\begin{array}{cr}
    u & -RC\dbar^T \\
    d & RC\ubar^T
    \end{array} \right)\ .
\label{ud}
\ee
Motivated by this arrangement we will refer to the left-multiplication
SU(2) as isospin symmetry, and to the right-multiplication SU(2)
as custodial symmetry.  It goes without saying that the two SU(2)'s
play a similar role, and the only ``preference'' for the left-multiplication
doublets is in our notation.  The isospin and custodial symmetries
get interchanged by $u \leftrightarrow -RC\ubar^T$, which is basically
the discrete symmetry~(\ref{eq:ccongR})
applied to the $u$ quark only.\footnote{
  We are free to add minus signs on the right-hand sides of
  Eqs.~(\ref{eq:ccongRa}) and~(\ref{eq:ccongRb}) simultaneously.
}

Let us take a closer look at the custodial-symmetry generator $\sigma_3$.
With reference to \Eq{lmult}, its action on the second row of $\xhat$, which is the multiplet
$(\phi_2,\phi_1^*)$, is $\d\phi_2=\phi_2$
and $\d\phi_1^* = -\phi_1^*$, or $\d\phi_1=\phi_1$.
Thus $\phi_1$ and $\phi_2$ transform with the same phase.
A translation to the language of \Eq{ud} is that the custodial $\sigma_3$
is just the baryon number.
In a two-flavor theory of complex-representation fermions,
the unbroken symmetries are isospin and the U(1) of baryon number.
In our case, the U(1) is enlarged
to a second SU(2) that we call the custodial symmetry, the
two other generators of which thus raise or lower the baryon number.

Now that we have understood the unbroken symmetry structure, let us
consider a few simple applications.  As a first exercise, one can show
that the transformation~(\ref{eq:ccongR}), when applied to the $u$ and $d$
fields simultaneously, is in fact an element of SO(4).
Indeed, consider $\Psi\to -i\sigma_2\Psi\, i\sigma_2$,
which is a simultaneous rotation in isospin and custodial SU(2).
This is just $u\to RC\ubar^T$, and the same for $d$.

We next turn to the NGBs.  Start with the familiar triplet of pions:
$\dbar \g_5 u$, $\ubar \g_5 d$, and $\ubar \g_5 u - \dbar \g_5 d$.
Now let us apply a custodial rotation of the form $\exp(i\theta\sigma_1)$.
Then $\ubar \g_5 d$ rotates into a linear combination of itself,
of $d^T RC \g_5 d$, and of $\ubar \g_5 RC \ubar^T$.
The last two are, respectively, a diquark and an antidiquark, each belonging to an isospin-1 multiplet.
It follows that there are indeed nine NGBs, which fall into three isospin triplets:
one made of diquarks, one of antidiquarks, and one of quark-antiquark pairs.

%%%%%%%%%%%%%%%%%%%%%%%%%%%%%%%%%%%%%%%%%%%%%%%%%%%%%%%%%%
\subsection{SU(4) $\leftrightarrow$ SO(6) correspondence
  and the six-quark baryon  \label{sec:B}}
%%%%%%%%%%%%%%%%%%%%%%%%%%%%%%%%%%%%%%%%%%%%%%%%%%%%%%%%%%

In this subsection we first work out in detail the identification
between the AS2 representation of SU(4) and the vector representation
of SO(6).  This allows us to construct a fully antisymmetric color
wave function for six AS2 fermions, which will be common to all our baryon
states.

In Sec.~\ref{sec:real} we labeled the components of the AS2 representation
by an index pair.  We can alternatively
introduce a single index $a=1,\ldots,6$, with the correspondence
$\psi_1=\psi_{[12]}$, $\psi_2=\psi_{[13]}$, $\ldots,$ $\psi_6=\psi_{[34]}$.
In the $\psi_a$ basis the matrix $R$
of Sec.~\ref{sec:real} takes the explicit form
\bee
  R = \left(\begin{array}{rrrrrr}
    0 &  0 &     0 &     0 &  0 & \ \ 1 \\
    0 &  0 &     0 &     0 & -1 &     0 \\
    0 &  0 &     0 & \ \ 1 &  0 &     0 \\
    0 &  0 & \ \ 1 &     0 &  0 &     0 \\
    0 & -1 &     0 &     0 &  0 &     0 \\
    1 &  0 &     0 &     0 &  0 &     0
    \end{array} \right)\ .
\label{R}
\ee
The inner product of two AS2 spinors,
\bee
  \sum_{i<j}\sum_{k<l} \e_{ijkl} \psi_{[ij]} \chi_{[kl]}
  = \psi_a R_{ab} \chi_b = \psi^T R \chi \ ,
\label{inner}
\ee
is SU(4) invariant by virtue of \Eq{eq:Uthru}.

We can recover the standard formulation
of SO(6) by applying a U(6) basis transformation to the AS2 states.
As a first step, we permute the basis elements and multiply
one of them by a minus sign, bringing $R$ to the block diagonal form
\bee
  R
  = \left(\begin{array}{ccc}
    \s_1 & 0 & 0 \\
    0 & \s_1 & 0 \\
    0 & 0 & \s_1
    \end{array} \right) \ .
\label{Rblock}
\ee
For a further change of basis, we note that
\bee
  \s_1 = \t^2, \ \text{with}\
  \t = \left(\begin{array}{cc}
      z & z^* \\
      z^* & z
      \end{array} \right) \ ,
\ee
where $z=(1+i)/{2}$.  We denote by $Q$ the $6\times6$ matrix with
three blocks of the matrix $\t$ along the main diagonal.
Upon performing the basis change
\bee
  \psi\to \psi' = Q\psi \ ,
\label{psi'}
\ee
the inner product becomes\footnote{
  The inner product is invariant
  under SU(4) [equivalently SO(6)] transformations, not under general
  U(6) transformations.
}
\bee
  \psi^T R \chi \to \psi^T Q^T R Q \chi = \psi^T \chi \ .
\ee
The inner product has now taken its standard SO(6) form.
Under the same basis change, the AS2 SU(4) generators transform as
\bee
  T_a \to Q T_a Q^\dagger\ .
\ee
Using the properties of the $R$ and $Q$ matrices and Eq.~(\ref{eq:Tthru})
it follows that
\bee
  (Q T_a Q^\dagger)^T = Q^\dagger T_a^T Q
  = Q R T_a^T R Q^\dagger = -Q T_a Q^\dagger \ .
\ee
In the new basis, the generators are antisymmetric
(and purely imaginary), as required for the standard basis of SO(6).

As an application of the above, we can show that the fully antisymmetric six-quark wave function
\bee
  B = \epsilon_{a_1a_2\cdots a_6} \psi_{a_1} \psi_{a_2}\cdots \psi_{a_6} \ ,
\label{B}
\ee
is gauge invariant (we suppress flavor indices).
To prove this, start from
\bee
  B' = \epsilon_{a_1a_2\cdots a_6} \psi'_{a_1} \psi'_{a_2}\cdots \psi'_{a_6} \ ,
\label{B'}
\ee
where the $\psi'$ basis was introduced in \Eq{psi'}.
This operator is clearly gauge invariant, because in the $\psi'$ basis
the SU(4) elements are mapped to orthogonal SO(6) matrices,
and the epsilon tensor
in \Eq{B'} is the invariant six-dimensional tensor.  Going back
to the original basis we have
\bee
  B' = \epsilon_{a_1a_2\cdots a_6} (Q\psi)_{a_1} (Q\psi)_{a_2}\cdots (Q\psi)_{a_6} \ .
\label{Bback}
\ee
The matrix $Q$ is unitary, and so it leaves invariant the epsilon tensor, up 
to a factor of $\det\, Q=-i$.  It follows that $B'=iB$, and hence $B$ is gauge invariant as well.
We use the fully antisymmetric color wave function (\ref{B})
in the construction of all baryon operators.

%%%%%%%%%%%%%%%%%%%%%%%%%%%%%%%%%%%%%%%%%%%%%%%%%%%%%%%%%%
\subsection{Diquarks, tetraquarks, and baryons \label{sec:tetra}}
%%%%%%%%%%%%%%%%%%%%%%%%%%%%%%%%%%%%%%%%%%%%%%%%%%%%%%%%%%
In constructing states with baryon number, we note first that a color-singlet state 
has to be made of an even number of quarks.
Thus we begin with diquarks.
As we have seen, the real color representation of the quarks leads to the conclusion 
that diquarks are degenerate with mesons.
Their color wave function involves the inner product~(\ref{inner})
\bee
D=\psi^f_a R_{ab}\psi^g_b,
\ee
where $f,g$ stand for the spin and flavor indices.
Because $R$ is symmetric, diquarks have a symmetric color wave function.
Viewed through the prism of the nonrelativistic quark model,
which puts the two quarks in an $s$-wave, the product of their spin and isospin wave
 functions must then be antisymmetric.%
\footnote{Since we are talking about diquarks rather than mesons, both quarks are 
in the $\sigma_3=+1$ state of the custodial SU(2).}
For (pseudo)scalars,
the spin wave function is antisymmetric,
and the isospin wave function should be symmetric.  Those that are NGBs have $I=1$, as seen above.
States with higher angular momentum are, of course, also possible.
These would include the diquark analogs of (axial) vector and tensor mesons.

The only way to construct a color-singlet {\em tetraquark\/} state is by pairwise 
contraction of the color indices,
\bee
T=(\psi^f_a R_{ab}\psi^g_b)(\psi^h_c R_{cd}\psi^i_d).
\ee
One can permute the spin-flavor indices to derive a total of three pairwise coupling schemes.
Linear combinations of these schemes will have mixed symmetry under color, but each 
term will still factor into two color-singlet diquarks.
Moreover, by applying an $RC$ transformation to one quark flavor at a time, one finds
 that the tetraquarks are degenerate with $\bar{\psi} \psi \psi \psi$ and 
$\bar{\psi} \bar{\psi} \psi \psi$ states.  It is an open question as to whether the
 tetraquark states in this theory will be meson and diquark scattering states, or 
bound states; we will not study them further here.

The first baryonic state that cannot be factored into smaller color-singlet components 
is the six-quark baryon written in \Eq{B}.
It differs essentially from the various pairwise contractions in that it is 
fully antisymmetric in color.
This makes it similar to the baryon of QCD, and indeed similar to baryons made of
 fundamental-representation quarks for any $N_c$.
Bolognesi \cite{Bolognesi:2006ws} has argued that this is the correct baryonic state
 for studying the large-$N_c$ limit of gauge theories with AS2 quarks.
In general, he finds that baryons made of $N_b=N_c(N_c-1)/2$ constituents in the AS2 
color representation fit well into a Skyrmion picture.
While our construction of the wave function~(\ref{B}) relies
on special properties of the $N_c=4$ theory, Bolognesi has given an existence proof
for a fully antisymmetric, gauge invariant color wave function for any $N_c$.

%%%%%%%%%%%%%%%%%%%%%%%%%%%%%%%%%%%%%%%%%%%%%%%%%%%%%%%%%%
\subsection{Interpolating fields \label{sec:interp}}
%%%%%%%%%%%%%%%%%%%%%%%%%%%%%%%%%%%%%%%%%%%%%%%%%%%%%%%%%%
A lattice simulation needs interpolating fields with an appropriate set
of quantum numbers.  As noted above, since it is fully antisymmetric under exchange,
the AS2 color wave function~(\ref{B}) is similar to the color wave function
of baryons made of fundamental representation fermions.  The multiplet patterns
are therefore similar as well.  The construction of baryon correlators
was discussed in detail in a previous work by one of us \cite{DeGrand:2012hd}.
Here we give a brief synopsis.

A convenient set of interpolating fields for baryons are operators
which create nonrelativistic quark model trial states. They are diagonal
in a $\gamma_0$ basis.  In the case at hand,
a generic two-flavor baryon interpolating field made out of $k$ up quarks
and $6-k$ down quarks can be written as
\bee
  O_B = \epsilon_{a_1 \cdots a_6} \,
  C^{s_1\cdots s_6}\, u_{a_1}^{s_1} \cdots  u_{a_k}^{s_k} \,
  d_{a_{k+1}}^{s_{k+1}} \cdots d_{a_6}^{s_6} \ ,
\label{eq:bar}
\ee
where summation over all color and spin indices is implied.
(We are free to put all the $u$'s to the left of all the $d$'s.)
The $C$'s are an appropriate set of Clebsch--Gordan coefficients.
The spin wave function of each quark type, $u$ or $d$,
must be totally symmetric.

Next we may take linear combinations of the $O_B$'s to construct
operators with definite isospin quantum numbers.  For states built of
two flavors of quarks all in the same spatial wave function,
multiplets are locked in equal values for angular momentum $J$ and isospin $I$.
Thus we have states with $I=J=3$, 2, 1, and 0.

The two-baryon correlator must include all nonzero contractions
of creation operators at the source and annihilation operators
at the sink. For each flavor, this gives a determinant of quark propagators.
These must be summed over all the ways that colors
can be apportioned between the quarks.
For the analog of the $\Delta^{++}$, the state with $I=I_3=J=J_3=N_b/2=3$,
this is a single term. The number of terms increases rapidly
as the angular momentum decreases,
raising the computational cost of the calculation.
(This was an issue for the $N_c=7$ baryons of Ref.~\cite{DeGrand:2012hd}.)
Fortunately,
$N_b=6$ is not too large and the calculation always remains manageable.

With baryon number as its third generator,
our baryons are highest-weight states of the custodial symmetry.
In this paper, we are content with studying these states,
and we do not consider the six-quark states with smaller baryon number
that would be needed to fill in multiplets of the custodial symmetry.

%%%%%%%%%%%%%%%%%%%%%%%%%%%%%%%%%%%%%%%%%%%%%%%%%%%%%%%%%%%%%%%%%%%%%
\section{Lattice action \label{sec:action}}
%%%%%%%%%%%%%%%%%%%%%%%%%%%%%%%%%%%%%%%%%%%%%%%%%%%%%%%%%%%%%%%%%%%%%

We define the lattice theory with the usual Wilson plaquette gauge action and with Wilson-clover fermions.
The fermion action uses gauge connections defined as normalized hypercubic (nHYP)
 smeared links~\cite{Hasenfratz:2001hp,Hasenfratz:2007rf,DeGrand:2012qa}.
The gauge coupling is set by the parameter $\beta = 2N_c / g_0^2$.  We take the two Dirac
 flavors to be degenerate, with common bare quark mass introduced via the hopping 
parameter $\kappa=(2m_0^{q} a+8)^{-1}$.
As is appropriate for nHYP smearing~\cite{Shamir:2010cq}, we fix the clover coefficient 
at its tree level value, $c_{\text{SW}}=1$.

nHYP smearing introduces a new type of discretization error, peculiar to the real 
representation of the matter field.
Our prescription for smearing the fermion's gauge connection begins with applying the
 nHYP formulas \cite{DeGrand:2012qa} to the fundamental gauge link, and then the resulting
 fat link $V_{ik}$ is promoted to the AS2 representation via \Eq{eq:recouple}.
The problem is that $V_{ik}$ is in fact an element of U($N_c$), not SU($N_c$), viz.,
\bee
V_{ik} = e^{i\theta} U_{ik},
\ee
where both the SU(4) part $U_{ik}$ and the U(1) phase $\theta$ are determined by our smearing recipe.
Having its origin in the smearing formulas, this U(1) phase
is a discretization effect, and hence it must vanish like some power of the lattice
spacing in the continuum limit.
When we apply Eq.~(\ref{eq:recouple})
to construct the AS2 fat link $V_{[ij][kl]}$ from the original fat link $V_{ik}$,
we end up with $V_{[ij][kl]} = e^{2i\theta} U_{[ij][kl]}$.
Because of this unphysical phase, the AS2 fat link $V_{[ij][kl]}$ fails
to satisfy the reality condition~(\ref{eq:Uthru}).
This in turn leads to violation of relations like \Eq{eq:relate}.

We can gauge the severity of this discretization error by looking at violations of \Eq{eq:relate}.
Comparing meson and diquark propagators calculated on single configurations, 
we have found differences in the third significant digit.
Similar effects are seen in the eigenvalue spectrum of the Wilson-clover operator.
To the extent that this error creeps into the generation of configurations, there is no cause for concern.

Nonetheless, the breaking of the symmetry (\ref{eq:Uthru}) in the observables is annoying.
A way to fix this problem is to replace the AS2 fat link $V_{[ij][kl]}$
obtained from \Eq{eq:recouple} with
\bee
V'_{[ij][kl]} = \frac{1}{2}(V + S V^* S)_{[ij][kl]}
\label{eq:correct}
\ee
before calculating observables.
(This can be regarded as a partial quenching, since the correction here is applied only to
 the valence fermions; one may use $V'$ for the sea fermions as well, but since we had
 already generated ensembles without this correction we chose not to do so.)

The new AS2 link $V'_{[ij][kl]}$ satisfies \Eq{eq:Uthru} by construction,
at the price of being slightly nonunitary.
We compared spectroscopy with and without this correction
for a $12^3 \times 24$ data set at one of our parameter
values ($\beta=9.6$, $\kappa=0.1285$).
The differences turned out to lie well under
one standard deviation.
We conclude that the discretization error and the partial quenching (\ref{eq:correct}) are benign.

Wilson fermions break chiral symmetry explicitly.
In the familiar case of a complex representation,
the symmetry-breaking pattern of the two-flavor continuum theory is
SU(2)$_L\times$SU(2)$_R$ $\to$ SU(2)$_V$.  With Wilson fermions,
the breaking of SU(2)$_L\times$SU(2)$_R$ becomes explicit,
and only SU(2)$_V$ (and, of course, baryon number) is a good symmetry.  While the NGBs become massless
when we tune $\kappa$ to its critical value $\kappa_c$, full chiral symmetry
is only restored in the continuum limit.  For $\kappa>\kappa_c$ one
enters the Aoki phase \cite{Aoki:1997fm,Aoki:2004iq},
where one of the NGB fields condenses, and SU(2)$_V$ is broken spontaneously.

In our model, the spontaneous symmetry breaking SU(4) $\to$ SO(4)
of a real representation turns into explicit breaking with Wilson fermions.
Only SO(4) is a good symmetry on the lattice, and the full SU(4) flavor
symmetry is only recovered in the continuum limit.
For $\kappa<\kappa_c$ the NGBs discussed in Sec.~\ref{sec:as2}---mesons and diquarks---acquire a mass.
For $\kappa>\kappa_c$ again
one expects to find an Aoki phase.  The case of five AS2 Majorana fermions
(relevant for the SU(5)/SO(5) nonlinear sigma model mentioned in the
Introduction) was recently studied using chiral Lagrangian techniques
in Ref.~\cite{Golterman:2014yha}.

%%%%%%%%%%%%%%%%%%%%%%%%%%%%%%%%%%%%%%%%%%%%%%%%%%%%%%%%%%%%%%%%%%%%%
\section{Phase diagram \label{sec:phase}}
\begin{figure}
\begin{center}
\includegraphics[width=\columnwidth,clip]{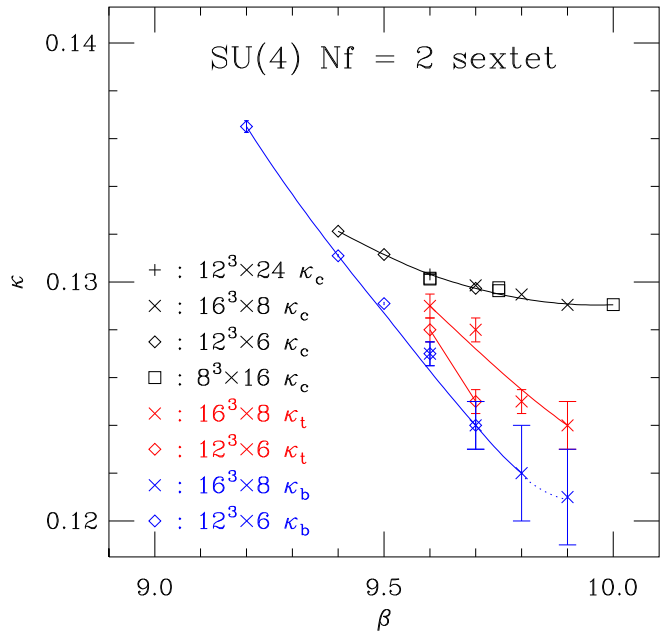}
\end{center}
\caption{Phase diagram of the SU(4) AS2 theory in the
($\beta$, $\kappa$) plane. The solid lines are drawn to guide the eye
and are not a fit to the data. From right to left: $\kappa_c$, $\kappa_t(N_t = 8)$,
$\kappa_t(N_t = 6)$, and $\kappa_b$.
The dotted line indicates weakening of the bulk transition to a crossover.
\label{fig:phase_diagram}}
\end{figure}

%%%%%%%%%%%%%%%%%%%%%%%%%%%%%%%%%%%%%%%%%%%%%%%%%%%%%%%%%%%%%%%%%%%%%
As preparation for spectroscopy, we have to find couplings in the confining and chirally broken phase.
The phase diagram of  Wilson fermion actions in the $(\beta, \kappa)$ plane can be
complicated, depending on the fermion content and the specific action
 used \cite{Aoki:1997fm,Aoki:2004iq, Golterman:2014yha, Farchioni:2004us, DeGrand:2010na}.
 Figure~\ref{fig:phase_diagram} shows the phase diagram we have observed for the SU(4) AS2 
action considered in this paper.
The curves shown indicate:
\begin{enumerate}
\item $\kappa_c(\beta)$, the critical value of the hopping parameter where the quark mass $m_q$ vanishes.
\item $\kappa_t(\beta)$, the curve of the thermal phase transition.
Its location shifts with the lattice size, and two lattice sizes are indicated.
\item $\kappa_b(\beta)$, the curve of a bulk phase transition that does not move with lattice size.
\end{enumerate}
We discuss each in turn.
%%%%%%%%%%%%%%%%%%%%%%%%%%%%%%%%%%%%%%%%%%%%%%%%%%%%%%%%%%%%%%%%%%%%
%%%%%%%%%%%%%%%%%%%%%%%%%%%%%%%%%%%%%%%%%%%%%%%%%%%%%%%%%%%%%%%%%%%%

%%%%%%%%%%%%%%%%%%%%%%%%%%%%%%%%%%%%%%%%%%%%%%%%%%%%%%%%%%%%%%%%%%%%
\subsection{$\kappa_c$ determination}
%%%%%%%%%%%%%%%%%%%%%%%%%%%%%%%%%%%%%%%%%%%%%%%%%%%%%%%%%%%%%%%%%%%%
We define the quark mass through the axial Ward identity (AWI), which relates the divergence of the
axial current $A_\mu^a=\bar \psi \gamma_\mu\gamma_5 (\tau^a/2)\psi$ to the pseudoscalar
density $P^a=\bar \psi \gamma_5 (\tau^a/2)\psi$.
At zero three-momentum we have
\bee
\partial_t \sum_\bx \svev{A_0^a(\bx,t)\co^a} = 2m_q \sum_\bx \svev{ P^a(\bx,t)\co^a}.
\label{eq:AWI}
\ee
where $\co^a$ is a source, here taken to
be a smeared ``Gaussian shell" source.
The critical $\kappa_c(\beta)$ line is determined through the vanishing of the
quark mass $m_q$.
As noted in Fig.~\ref{fig:phase_diagram}, we use several lattice sizes $N_s^3\times N_t$.
When $N_t>N_s$, $t$ labels the usual temporal direction, but when $N_t<N_s$, we choose one 
of the spatial directions to be $t$ in \Eq{eq:AWI}
from correlators taken along one of the spacial directions of the lattice (so that the sum over
 $x$ in Eq.~\ref{eq:AWI} includes two directions with periodic fermion boundary 
conditions and one antiperiodic direction).

\begin{figure*}[b]
\begin{center}
\includegraphics[width=\textwidth,clip]{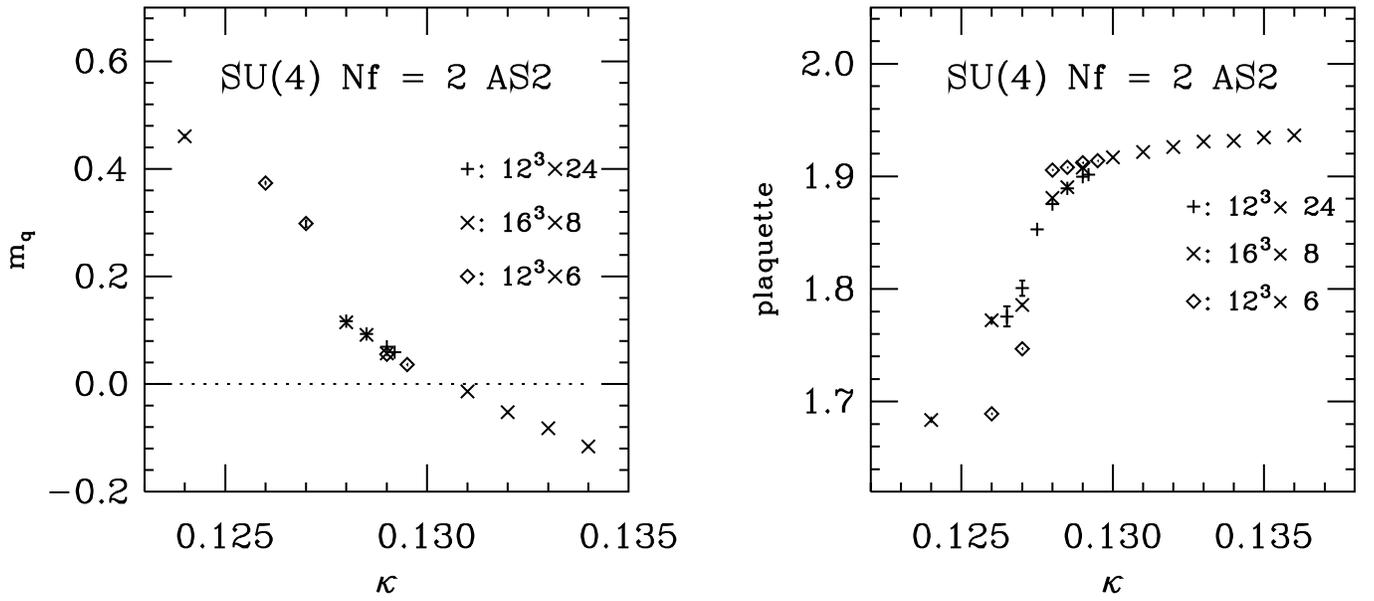}
\end{center}
\caption{
Left: Quark mass $m_q$ as a function of $\kappa$ in different volumes at $\beta = 9.6$.
Right: Average plaquette as a function of $\kappa$ in different volumes at $\beta = 9.6$.
\label{fig:bulk}}
\end{figure*}

One example of the $\kappa$
dependence of the quark mass $m_q$ is shown in the left panel of Fig.~\ref{fig:bulk}.
%%%%%%%%%%%%%%%%%%%%%%%%%%%%%%%%%%%%%%%%%%%%%%%%%%%%%%%%%%%%%%%%%%%%%
%%%%%%%%%%%%%%%%%%%%%%%%%%%%%%%%%%%%%%%%%%%%%%%%%%%%%%%%%%%%%%%%%%%%%
The zero crossing at $\kappa_c$ is apparent, as is a discontinuity in the $m_q(\kappa)$.
The latter is a signal of a bulk transition, which we discuss below in Sec.~\ref{subsec:bulk}.
We note that there is little volume dependence in $\kappa_c(\beta)$.
We plot $\kappa_c(\beta)$ from all the volumes in Fig. \ref{fig:phase_diagram} as a black line.

%%%%%%%%%%%%%%%%%%%%%%%%%%%%%%%%%%%%%%%%%%%%%%%%%%%%%%%%%%%%%%%%%%%%
\subsection{$\kappa_t$ determination}
\begin{figure*}
\begin{center}
\includegraphics[width=1.0\textwidth,clip]{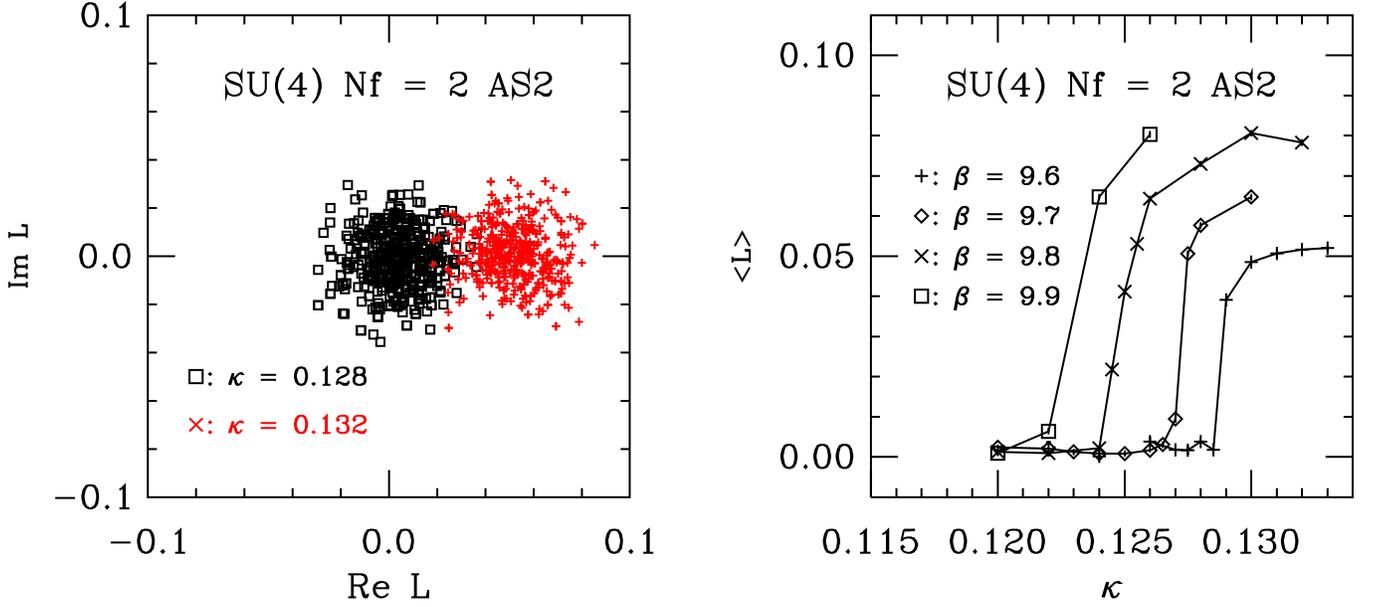}
\end{center}
\caption{
Left: Scatter plots of the Polyakov loop in the two different phases
on  $16^3\times 8$ lattices at $\beta = 9.6$.
Right: Average Polyakov loop on $16^3\times 8$ lattices for different $\beta$ and
 $\kappa$ values. The jump of the average Polyakov loop values for each $\beta$
value signals a finite-temperature transition.
\label{fig:finiteT}}
\end{figure*}

%%%%%%%%%%%%%%%%%%%%%%%%%%%%%%%%%%%%%%%%%%%%%%%%%%%%%%%%%%%%%%%%%%%%
The finite-temperature transition lines $\kappa_t(\beta)$ are determined from the
behavior of the Polyakov loop $L$.
With AS2 fermions, the $Z_4$ center symmetry of the pure-gauge theory
symmetry is broken only to $Z_2$ and therefore there is a true confinement phase transition in our theory.
$\langle L\rangle=0$ in the low-temperature phase, while in the high-temperature phase, 
 $\langle L\rangle$ orders along the real axis.
Typical scatter plots of the Polyakov loop in the two phases are shown in the left panel
 of Fig.~\ref{fig:finiteT}.
The average Polyakov loop as a function of $\kappa$ at four different $\beta$
values for a  $16^3\times8$ volume are shown in the right panel of Fig.~\ref{fig:finiteT}.
%%%%%%%%%%%%%%%%%%%%%%%%%%%%%%%%%%%%%%%%%%%%%%%%%%%%%%%%%%%%%%%%%%%%%
%%%%%%%%%%%%%%%%%%%%%%%%%%%%%%%%%%%%%%%%%%%%%%%%%%%%%%%%%%%%%%%%%%%%%

The $\kappa_t(\beta)$ lines for two different volumes, $16^3\times 8$ and $12^3\times 6$,
 are shown in Fig.~\ref{fig:phase_diagram} as red lines. The transition moves to weaker 
coupling as $N_t$ increases, as expected from asymptotic freedom.

%%%%%%%%%%%%%%%%%%%%%%%%%%%%%%%%%%%%%%%%%%%%%%%%%%%%%%%%%%%%%%%%%%%%%
\subsection{$\kappa_b$ determination} \label{subsec:bulk}
%%%%%%%%%%%%%%%%%%%%%%%%%%%%%%%%%%%%%%%%%%%%%%%%%%%%%%%%%%%%%%%%%%%%%
In addition to the temperature-dependent deconfinement lines $\kappa_t(\beta)$, our system 
exhibits another transition line $\kappa_b(\beta)$.
Its presence is signaled by discontinuities in several observables, notably the average 
plaquette and the quark mass $m_q$.
We find that the position of the discontinuity is independent of volume.
This is a bulk transition associated with the particular lattice action we use; most likely
 it has nothing to do with continuum physics.
The mechanism that triggers the bulk transition is not clear to us.
Similar behavior has been observed in other lattice actions, when the number of
fermionic degrees of freedom is large~\cite{Aoki:1997fm,Aoki:2004iq, Farchioni:2004us, DeGrand:2010na}.
  A similar bulk transition has been observed in studies of the SU(4) pure-gauge theory 
at $\beta \sim 10.2$ \cite{Barkai:1982kp}.

We have already seen, in the left panel of Fig.~\ref{fig:bulk}, a discontinuity in 
 the quark mass $m_q$ at $\kappa \approx 0.127$ as we scan at $\beta=9.6$.
The right panel of Fig.~\ref{fig:bulk}  shows the average plaquette values
for $\beta = 9.6$ in three different volumes.
All the plaquette data show a sudden jump at the same value of $\kappa$.

The transition weakens in the large-$\beta$/small-$\kappa$ region and appears  
to show only a smooth crossover at $\beta \approx 10.0$.
The $\kappa_b(\beta)$ line, determined on two different volumes, $16^3\times 8$ 
and $12^3\times 6$,  is sketched in
Fig. \ref{fig:phase_diagram} in blue.
Further work is needed to understand the  origin of this peculiar bulk transition.
For the current study, however, we only need to make sure that our
simulation is on the weak-coupling (large-$\beta$) side of this transition so that
it has a direct connection to continuum physics.

%%%%%%%%%%%%%%%%%%%%%%%%%%%%%%%%%%%%%%%%%%%%%%%%%%%%%%%%%%%%%%%%%%%%%
\section{Spectroscopy \label{sec:spectra}}
%%%%%%%%%%%%%%%%%%%%%%%%%%%%%%%%%%%%%%%%%%%%%%%%%%%%%%%%%%%%%%%%%%%%%

%%%%%%%%%%%%%%%%%%%%%%%%%%%%%%%%%%%%%%%%%%%%%%%%%%%%%%%%%%%%%%%%%%%%%
\subsection{SU(4) AS2 \label{sec:spectra4}}
%%%%%%%%%%%%%%%%%%%%%%%%%%%%%%%%%%%%%%%%%%%%%%%%%%%%%%%%%%%%%%%%%%%%%
Referring to the phase diagram, Fig.~\ref{fig:phase_diagram}, we chose to simulate
 the SU(4) AS2 theory at $\beta=9.6$ for a range of hopping parameter 
values $0.127<\kappa<0.130$, between the bulk transition and $\kappa_c$.
Our simulation volumes were all $16^3\times 32$, and the resulting spectra show 
that our chosen $\kappa$ values kept us in the confining phase, $\kappa<\kappa_t$. 
 Gauge-field updates used the HMC algorithm with a multilevel Omelyan integrator, 
including one level of mass preconditioning for the fermions; integration parameters 
were adjusted to maintain acceptance rates on the order of 70\%-80\%.  Gauge
 configurations were saved to disk every ten updates.  The simulations are 
summarized in Table~\ref{tab:SU4par}.

%%%%%%%%%%%%%%%%%%%%%%%%%%%%%%%%%%%%%%%%%%%%%%%%%%%%%%%%%%%%%%%%%%%%%
\begin{table}
\begin{ruledtabular}
\begin{tabular}{lcc}
$\kappa$&Configurations&$r_1/a$\\
\hline
0.128 & 146 & 2.50(1) \\
0.1285& 140 & 2.78(2) \\
0.129 & 200 & 2.97(2) \\
0.1292& 161 & 3.22(3)
\end{tabular}
\end{ruledtabular}
\caption{Parameters of the SU(4) AS2 simulations.  All are at coupling $\beta=9.6$, in volume $16^3\times32$.
\label{tab:SU4par}}
\end{table}
%%%%%%%%%%%%%%%%%%%%%%%%%%%%%%%%%%%%%%%%%%%%%%%%%%%%%%%%%%%%%%%%%%%%%

\begin{figure*}
\begin{center}
\includegraphics[width=\textwidth,clip]{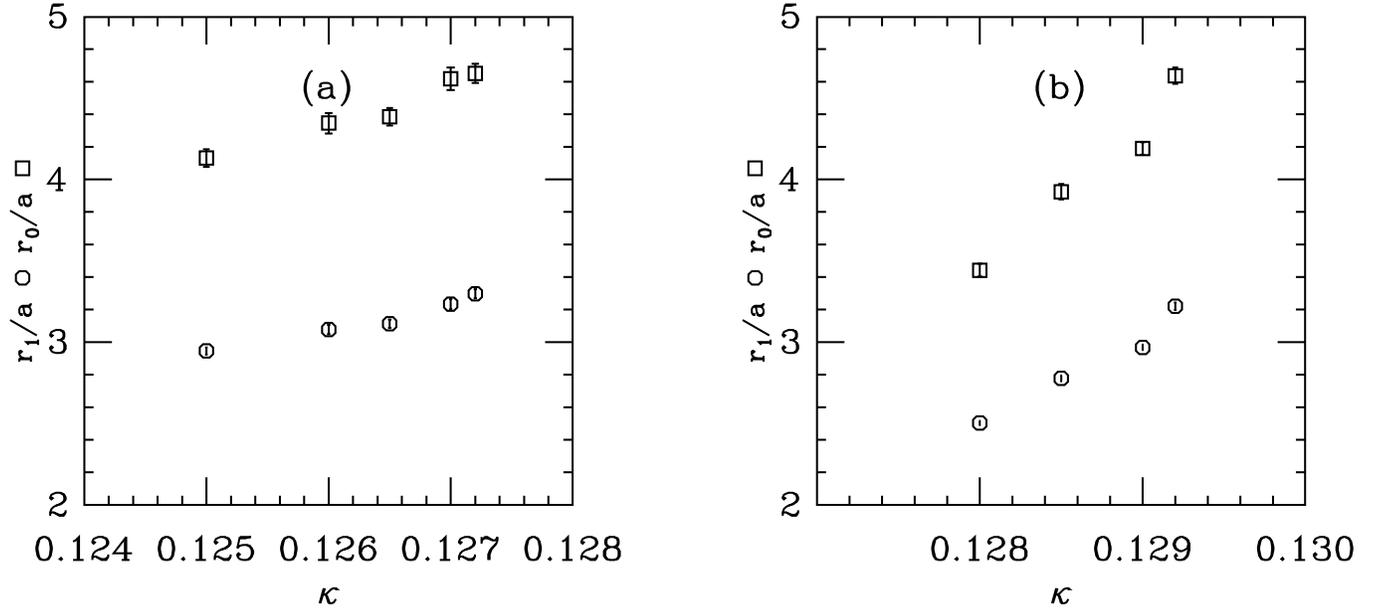}
\end{center}
\caption{Sommer parameters $r_0$ and $r_1$ from the dynamical SU(3) and SU(4) data 
sets [panels (a) and (b), respectively].
\label{fig:r0r1}}
\end{figure*}

The coupling $\beta=9.6$ gives a lattice spacing that is neither too large nor too small.
For comparison with other theories, we fix the lattice spacing using
the shorter version \cite{Bernard:2000gd} of the Sommer~\cite{Sommer:1993ce} parameter
$r_1$, defined in terms of the force $F(r)$ between static quarks:
$r^2 F(r)= -1.0$ at $r=r_1$.
The real-world value is $r_1= 0.31$~fm~\cite{Bazavov:2009bb},
and thus Table~\ref{tab:SU4par} shows that our lattice spacings would correspond
 to a length scale of approximately 0.1~fm in QCD.
For later comparison, we plot both Sommer parameters for our simulations in Fig.~\ref{fig:r0r1}.
%%%%%%%%%%%%%%%%%%%%%%%%%%%%%%%%%%%%%%%%%%%%%%%%%%%%%%%%%%%%%%%%%%%%%
%%%%%%%%%%%%%%%%%%%%%%%%%%%%%%%%%%%%%%%%%%%%%%%%%%%%%%%%%%%%%%%%%%%%%

In addition to the simulations listed in Table~\ref{tab:SU4par}, we used the 
$\kappa=0.129$ lattices as a set of configurations on which we computed
partially quenched (PQ) spectroscopy with four values of the valence quark mass,
$\kappa_V=0.1295$, 0.130, 0.1305, and 0.131. These data sets used the full complement of
$\kappa=0.129$ configurations. Of course, their lattice spacing is the same as that of the $\kappa=0.129$ set.

The correlation functions of which the analysis produced our spectroscopy used
propagators constructed in the Coulomb gauge, the sources of which were Gaussians.
We used $\vec p=0$ point sinks.
We collected sets for several different values of the width $R_0$ of the source.
These correlation functions are not variational since the source and sink are different.
We begin each fit with a distance-dependent effective mass $\meff(t)$, defined to be 
$\meff(t) = \log C(t)/C(t+1)$ consistent with open boundary conditions for the
 correlator $C(t)$. Because our sources and sinks are not identical,
$\meff(t)$ can approach its asymptotic value from above or below.
We mixed data with different values of $R_0$ to produce correlators with relatively
 flat $\meff(t)$, which we then used in a full analysis involving fits to a wide range of $t$'s.
For more detail see Ref.~\cite{DeGrand:2012hd}.

Our resulting data are shown in Tables~\ref{tab:su4}, \ref{tab:su4bar}, and~\ref{tab:su4bardiff}.
Table~\ref{tab:su4} shows the AWI mass and meson masses and decay constants.
The pseudoscalar and vector meson decay constants, the definitions of which are given in
 Eqs.~(\ref{eq:fpi}) and~(\ref{eq:fv}) below, are given with lattice normalization 
for the fermion fields. The conversion to continuum numbers will be described below.

We also measured the masses of the $J=0$ and $J=1$ diquarks using nonrelativistic
 quark model interpolating fields, diquark analogs of the operators we used for baryons.
Their masses are, as expected, degenerate with those of their mesonic partners.

Tables~\ref{tab:su4bar} and~\ref{tab:su4bardiff} give the baryon masses and mass differences.
These are computed together: a jackknife average of correlated, single-exponential
 fits to all four masses is performed and the differences are collected.
This insures that the average mass difference is indeed the difference of the average masses. Since the data
sets for the different angular-momentum states are identical, the uncertainty in the mass difference is
usually smaller than the naive combination of uncertainties on the individual masses. These fits
are over the range $5\le t \le 10$. We have checked that fits over nearby $t$ ranges are consistent
within uncertainties with these results.
We omit results for $\kappa_V=0.131$ because the uncertainties in the baryon masses,
especially the $J=0$ baryon, are very large.

%%%%%%%%%%%%%%%%%%%%%%%%%%%%%%%%%%%%%%%%%%%%%%%%%%%%%%%%%%%%%%%%%%%%%
\begin{table}
\begin{ruledtabular}
\begin{tabular}{l c c c ll }
$\kappa$  & $am_q$ & $am_{PS}$ &   $am_V$ &  $af_{PS}$ &  $af_V$ \\
\hline
 0.1280 & 0.124 &  0.680(1) & 0.888(3) & 0.978(10)  & 1.559(18) \\
 0.1285 & 0.089 &  0.554(2) & 0.749(6) & 0.730(12)  & 1.563(29) \\
 0.1290 & 0.067 &  0.462(1) & 0.666(3) & 0.693(5)  & 1.516(4) \\
 0.1292 & 0.057 &  0.417(2) & 0.602(2) & 0.637(7)  & 1.501(7) \\[2pt]
 0.1295\footnotemark[1] & 0.053 &  0.409(1) & 0.630(2) & 0.654(3)  & 1.570(8) \\
 0.1300\footnotemark[1] & 0.039 &  0.350(2) & 0.596(3) & 0.627(3)  & 1.627(11) \\
 0.1305\footnotemark[1] & 0.025 &  0.281(2) & 0.561(4) & 0.596(4)  & 1.699(13) \\
 0.1310\footnotemark[1] & 0.011 &  0.190(4) & 0.529(8) & 0.562(5)  & 1.783(18) \\
 \end{tabular}
\end{ruledtabular}
\footnotetext[1]{Partially quenched: same gauge configurations as $\kappa=0.129$.}
\caption{AWI mass and meson spectra and decay constants from dynamical SU(4) AS2 simulations.
$f_{PS}$ and $f_V$ have lattice normalization.
\label{tab:su4}}
\end{table}
%%%%%%%%%%%%%%%%%%%%%%%%%%%%%%%%%%%%%%%%%%%%%%%%%%%%%%%%%%%%%%%%%%%%%

%%%%%%%%%%%%%%%%%%%%%%%%%%%%%%%%%%%%%%%%%%%%%%%%%%%%%%%%%%%%%%%%%%%%%
\begin{table}
\begin{ruledtabular}
\begin{tabular}{l c c c c}
$\kappa$  & $aM_B(3)$ & $aM_B(2)$ & $aM_B(1)$ & $aM_B(0)$  \\
\hline
0.1280 &  3.134(43)& 3.055(32)  & 2.972(30)& 2.923(32)\\
0.1285 &  2.608(35)& 2.513(27)  & 2.442(25)& 2.389(23)\\
0.1290 &  2.297(21)& 2.212(17)  & 2.147(16)& 2.113(16)\\
0.1292 &  2.046(24)& 1.990(22)  & 1.948(22)& 1.920(18)\\[2pt]
0.1295\footnotemark[1] &  2.179(26)& 2.075(20)& 2.002(20)& 1.972(18) \\
0.1300\footnotemark[1] &  2.094(37)& 1.964(35)& 1.902(29)& 1.848(28) \\
0.1305\footnotemark[1] &  1.984(77)& 1.854(52)& 1.826(95)& 1.732(57) \\
\end{tabular}
\end{ruledtabular}
\footnotetext[1]{Partially quenched: same gauge configurations as $\kappa=0.129$.}
\caption{Baryon masses from dynamical SU(4) AS2 simulations.
The number labels the angular momentum of the state: $M_B(3)=M_B(J=3)$.
\label{tab:su4bar}}
\end{table}
%%%%%%%%%%%%%%%%%%%%%%%%%%%%%%%%%%%%%%%%%%%%%%%%%%%%%%%%%%%%%%%%%%%%%

%%%%%%%%%%%%%%%%%%%%%%%%%%%%%%%%%%%%%%%%%%%%%%%%%%%%%%%%%%%%%%%%%%%%%
\begin{table}
\begin{ruledtabular}
\begin{tabular}{llll}
$\kappa$  & $a\Delta M_{23}$ &  $a\Delta M_{13}$ & $a\Delta M_{03}$ \\
\hline
0.1280 & 0.079(29)& 0.162(34)& 0.210(34) \\
0.1285 & 0.095(26)& 0.166(33)& 0.219(34) \\
0.1290 & 0.086(16)& 0.151(17)& 0.185(19) \\
0.1292 & 0.056(16)& 0.098(22)& 0.126(12) \\[2pt]
0.1295\footnotemark[1] & 0.104(21)& 0.177(22)& 0.207(23) \\
0.1300\footnotemark[1] & 0.131(36)& 0.192(37)& 0.247(37) \\
0.1305\footnotemark[1] & 0.131(66)& 0.159(117)& 0.253(80) \\
 \end{tabular}
\end{ruledtabular}
\footnotetext[1]{Partially quenched: same gauge configurations as $\kappa=0.129$.}
\caption{Baryon  mass splittings from dynamical SU(4) AS2 simulations.
We define $\Delta M_{J_1J_2}\equiv M_B(J_2)-M_B(J_1)$.
\label{tab:su4bardiff}}
\end{table}
%%%%%%%%%%%%%%%%%%%%%%%%%%%%%%%%%%%%%%%%%%%%%%%%%%%%%%%%%%%%%%%%%%%%%

%%%%%%%%%%%%%%%%%%%%%%%%%%%%%%%%%%%%%%%%%%%%%%%%%%%%%%%%%%%%%%%%%%%%%
\subsection{SU(3) fundamental \label{sec:spectra3}}
%%%%%%%%%%%%%%%%%%%%%%%%%%%%%%%%%%%%%%%%%%%%%%%%%%%%%%%%%%%%%%%%%%%%%
We also generated a data set for SU(3) gauge fields coupled to $N_f=2$ fermions in the 
fundamental representation.
We did this for two (related) reasons.
First, the SU(4) data sets include dynamical fermions,
and so we felt that our comparison to $N_c=3$ ought to be dynamical to dynamical.
Second, all previous large-$N_c$ comparisons were of quenched data sets.
While quenching is not the state of the art,
at the quark masses at which we work one might expect quenching artifacts to be small.
A direct comparison seemed to be in order, and we found (as expected) that quenching effects indeed were small.
This will be seen in the figures.
 Again, we used the clover action with nHYP links and $c_{SW}=1$.
The lattice volume was again $16^3\times 32 $ sites.
 We chose a gauge coupling $\beta=5.4$.  We saved configurations every five HMC trajectories. Parameter values
are shown in Table \ref{tab:SU3par}.
Table~\ref{tab:su3} shows mesonic observables from the dynamical SU(3) simulations,
while Table~\ref{tab:su3bar}  shows  baryon masses and mass differences. These
numbers are taken from a jackknife average of the data sets with a fit range $5\le t\le 10$.

%%%%%%%%%%%%%%%%%%%%%%%%%%%%%%%%%%%%%%%%%%%%%%%%%%%%%%%%%%%%%%%%%%%%%
\begin{figure*}[h]
\begin{center}
\includegraphics[width=\textwidth,clip]{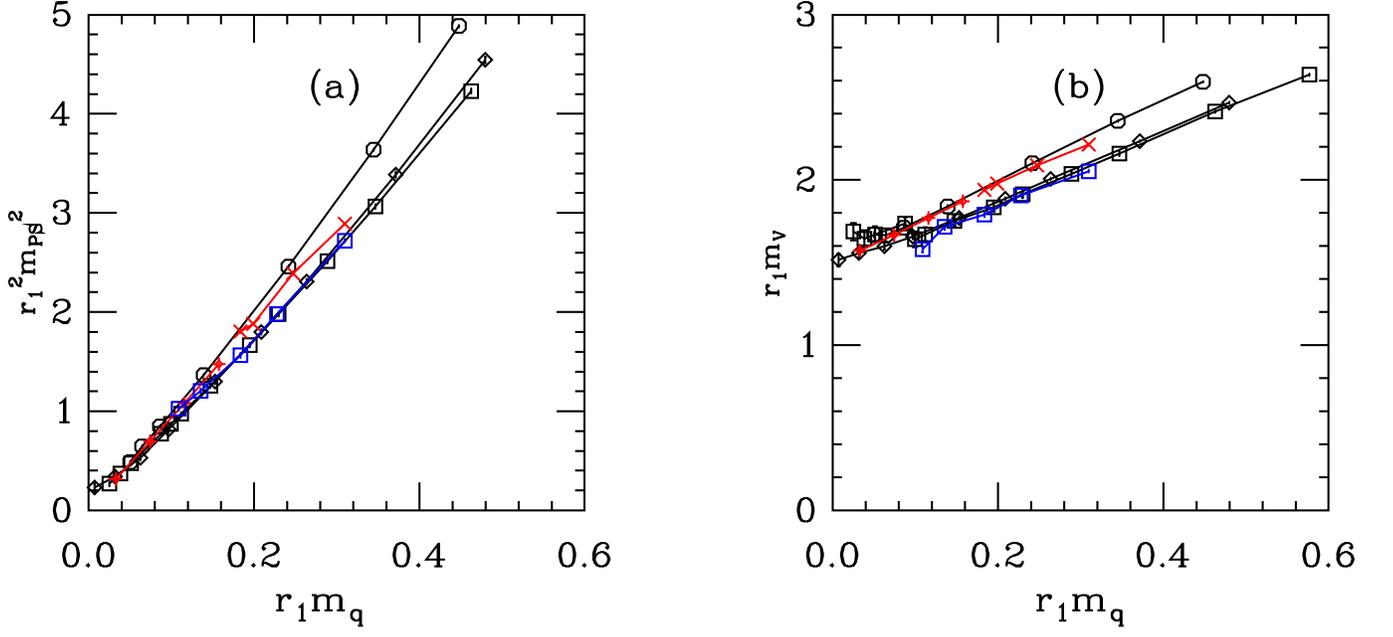}
\end{center}
\caption{Meson spectroscopy. On the left, the squared pseudoscalar mass scaled by  $r_1^2$, on the right,
$r_1$ times the vector meson mass. The abscissa is $r_1$ times the AWI quark mass.
The data sets are:
black squares for quenched SU(3) fundamentals, black diamonds for quenched SU(5) fundamentals,
black octagons for quenched SU(7) fundamentals, and 
red crosses for SU(4) AS2; the fancy diamonds are the PQ data. Finally, the blue squares
are  SU(3) with two dynamical, fundamental flavors.
\label{fig:mesons}}
\end{figure*}
%%%%%%%%%%%%%%%%%%%%%%%%%%%%%%%%%%%%%%%%%%%%%%%%%%%%%%%%%%%%%%%%%%%%%

%%%%%%%%%%%%%%%%%%%%%%%%%%%%%%%%%%%%%%%%%%%%%%%%%%%%%%%%%%%%%%%%%%%%%
\begin{figure*}[h]
\begin{center}
\includegraphics[width=\textwidth,clip]{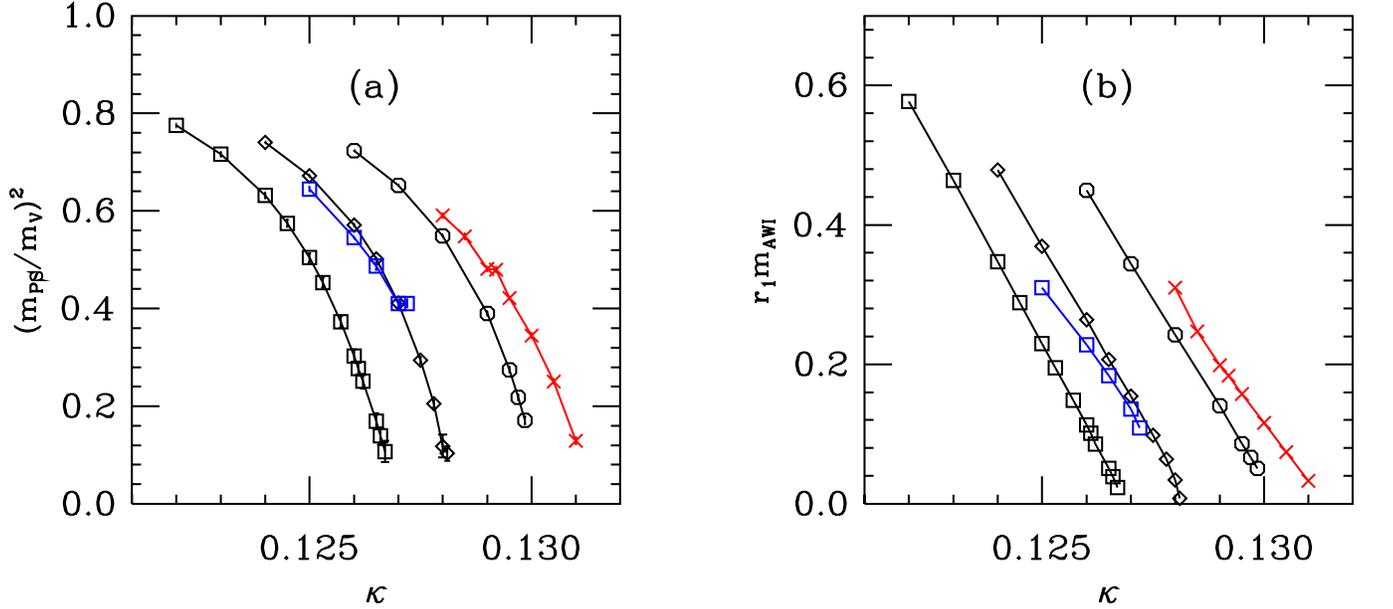}
\end{center}
\caption{Two ways to match bare parameters: panel (a) $(m_{PS}/m_V)^2$ vs $\kappa$, and panel (b)
$r_1 m_{AWI}$ vs $\kappa$.
The data sets are:
black squares for quenched SU(3) fundamentals, black diamonds for quenched SU(5) fundamentals,
black octagons for quenched SU(7) fundamentals, and
red crosses for SU(4) AS2. Finally, the blue squares
are  SU(3) with two dynamical, fundamental flavors.
\label{fig:match}}
\end{figure*}
%%%%%%%%%%%%%%%%%%%%%%%%%%%%%%%%%%%%%%%%%%%%%%%%%%%%%%%%%%%%%%%%%%%%%

%%%%%%%%%%%%%%%%%%%%%%%%%%%%%%%%%%%%%%%%%%%%%%%%%%%%%%%%%%%%%%%%%%%%%
\begin{table}
\begin{ruledtabular}
\begin{tabular}{lcc}
$\kappa$&Configurations&$r_1/a$\\
\hline
0.125 &100& 2.95(2)\\
0.126 &100& 3.08(3)\\
0.1265&100& 3.11(3)\\
0.127 &100& 3.23(3)\\
0.1272&100& 3.30(3)
\end{tabular}
\end{ruledtabular}
\caption{Parameters of the SU(3) simulations. All are at coupling $\beta=5.4$, in volume $16^3\times32$.
\label{tab:SU3par}}
\end{table}
%%%%%%%%%%%%%%%%%%%%%%%%%%%%%%%%%%%%%%%%%%%%%%%%%%%%%%%%%%%%%%%%%%%%%

%%%%%%%%%%%%%%%%%%%%%%%%%%%%%%%%%%%%%%%%%%%%%%%%%%%%%%%%%%%%%%%%%%%%%
\begin{table}
\begin{ruledtabular}
\begin{tabular}{c c c c c l}
$\kappa$  & $am_q$ & $am_{PS}$ & $am_V$ &  $af_{PS}$ & $af_V$ \\
\hline
0.1250 & 0.105 &  0.559(2)& 0.696(3)& 0.456(6) & 0.905(4) \\
0.1260 & 0.070 &  0.457(1)& 0.619(3)& 0.424(4) & 0.993(8) \\
0.1265 & 0.059 &  0.402(3)& 0.576(5)& 0.385(3) & 1.001(9) \\
0.1270 & 0.042 &  0.340(3)& 0.531(5)& 0.370(5) & 1.050(9) \\
0.1272 & 0.028 &  0.307(3)& 0.479(6)& 0.318(7) & 1.037(13)\\
 \end{tabular}
\end{ruledtabular}
\caption{AWI mass and meson spectra and decay constants from dynamical SU(3) simulations.
$f_{PS}$ and $f_V$ have lattice normalization.
\label{tab:su3}}
\end{table}
%%%%%%%%%%%%%%%%%%%%%%%%%%%%%%%%%%%%%%%%%%%%%%%%%%%%%%%%%%%%%%%%%%%%%

%%%%%%%%%%%%%%%%%%%%%%%%%%%%%%%%%%%%%%%%%%%%%%%%%%%%%%%%%%%%%%%%%%%%%
\begin{table}
\begin{ruledtabular}
\begin{tabular}{clll}
$\kappa$  &  $aM_B(3/2)$  & $aM_B(1/2)$ & $a\Delta M$ \\
\hline
0.1250 & 1.143(13)& 1.042(7)& 0.100(11)\\
0.1260 & 1.011(10)& 0.926(7)& 0.085(9)\\
0.1265 & 0.959(18)& 0.838(11)& 0.120(16)\\
0.1270 & 0.887(23)& 0.748(8)& 0.139(22)\\
0.1272 & 0.833(25)& 0.698(8)& 0.135(24)\\
 \end{tabular}
\end{ruledtabular}
\caption{Baryon masses and splittings from dynamical SU(3) simulations.
The number labels the angular momentum of the state: $M_B(1/2)=M_B(J=1/2)$.
The difference is $\Delta M\equiv M_B(3/2) - M_B(1/2)$.
\label{tab:su3bar}}
\end{table}
%%%%%%%%%%%%%%%%%%%%%%%%%%%%%%%%%%%%%%%%%%%%%%%%%%%%%%%%%%%%%%%%%%%%%

%%%%%%%%%%%%%%%%%%%%%%%%%%%%%%%%%%%%%%%%%%%%%%%%%%%%%%%%%%%%%%%%%%%%%
\section{Comparisons: Mesons \label{sec:mesons}}
%%%%%%%%%%%%%%%%%%%%%%%%%%%%%%%%%%%%%%%%%%%%%%%%%%%%%%%%%%%%%%%%%%%%%

We have presented our results for meson and baryon spectra and also for meson decay 
constants in the SU(4) AS2 and SU(3) theories in Sec.~\ref{sec:spectra}.
In this section and the next we will plot and rescale them for comparison with each 
other and with the quenched SU($N_c$) theories, $N_c=3$, 5, and~7.  This is an important consistency check on our results, and discrepancies with predicted scaling may point to interesting directions for future study of AS2 theories with large $N_c$.

%%%%%%%%%%%%%%%%%%%%%%%%%%%%%%%%%%%%%%%%%%%%%%%%%%%%%%%%%%%%%%%%%%%%%
\subsection{Spectroscopy}
%%%%%%%%%%%%%%%%%%%%%%%%%%%%%%%%%%%%%%%%%%%%%%%%%%%%%%%%%%%%%%%%%%%%%

We plot the data for the pseudoscalar and vector meson masses in Fig.~\ref{fig:mesons}.
To set the scale, we use the Sommer parameter $r_1$, and for the quark mass we use the
lattice-regulated  AWI quark mass, scaled  by $r_1$ in the plots.

We display several data sets together. The new ones are the SU(4) AS2 sets, shown in
 red (crosses for the full dynamical sets and fancy diamonds for the partially quenched ones),
 and the dynamical SU(3) sets (blue squares).
The black squares, diamonds, and octagons are previously published data from quenched
 simulations with $N_c=3$, 5, and~7 with fundamental fermions \cite{DeGrand:2012hd}.

To carry out meaningful comparisons between data obtained at different $N_c$'s, we must match the bare parameters between the simulations in some common way.  This is an inherently ambiguous procedure, but let us make the attempt.
We know that hadron masses depend monotonically on the quark mass.
We can compare results at the same values of the quark mass by selecting data at constant 
$(m_{PS}/m_V)^2$---this is a quantity which is roughly linear in the quark mass---or we can
 use the AWI quark mass itself, rendered dimensionless by multiplication by $r_1$.
These comparisons are shown in Fig.~\ref{fig:match}.
For both quantities, the theories can be matched over almost the entire range of $\kappa$.

\begin{figure}
\begin{center}
\includegraphics[width=\columnwidth,clip]{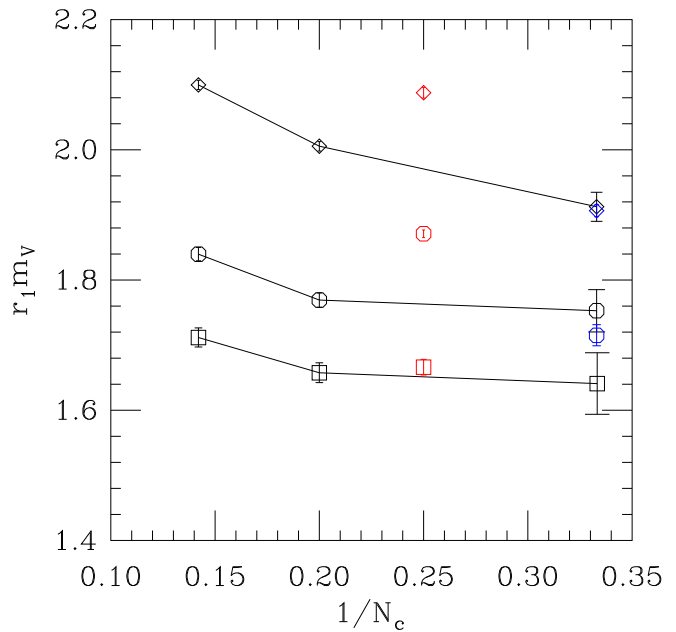}
\end{center}
\caption{
Variation of $r_1 m_V$ vs $1/N_c$ for roughly matched data using $(m_{PS}/m_V)^2$.
The  diamonds are for $(m_{PS}/m_V)^2 = 0.54-0.56$, octagons for  $(m_{PS}/m_V)^2 = 0.40$,
and squares for $(m_{PS}/m_V)^2 = 0.29-0.32$. The blue symbols are the dynamical SU(3) data,
 and the red symbols are the
SU(4) AS2 data. Black symbols show quenched fundamental results.
\label{fig:vectorsvs1n}}
\end{figure}

We now select matching points for which we have many data sets.
Thus we choose to use $(m_{PS}/m_V)^2 = 0.54$--0.56, 0.40, and 0.29--0.32 as the three ratios.
We plot $r_1 m_V$ as a function of $1/N_c$, since we expect the leading corrections to scale with $1/N_c$.
The result is shown in Fig.~\ref{fig:vectorsvs1n}.
%%%%%%%%%%%%%%%%%%%%%%%%%%%%%%%%%%%%%%%%%%%%%%%%%%%%%%%%%%%%%%%%%%%%%
%%%%%%%%%%%%%%%%%%%%%%%%%%%%%%%%%%%%%%%%%%%%%%%%%%%%%%%%%%%%%%%%%%%%%

To leading order in the expansion, meson masses in both the fundamental and AS2 theories are expected to be independent of $N_c$ \cite{Cherman:2006iy}.  Empirically, it appears that the systems connected by the original 't Hooft large-$N_c$ scaling
 argument---fundamental fermions---show smaller $1/N_c$ variation than the AS2  
systems over the range of $N_c$ shown.
In particular, we note that the AS2 data with $N_c=4$ and the fundamental data with $N_c=7$ show roughly the same shift compared to $N_c=3$.  This is seen to be the case for all quark mass values (see Fig.~\ref{fig:mesons}).

%%%%%%%%%%%%%%%%%%%%%%%%%%%%%%%%%%%%%%%%%%%%%%%%%%%%%%%%%%%%%%%%%%%%%
\subsection{Decay constants}
%%%%%%%%%%%%%%%%%%%%%%%%%%%%%%%%%%%%%%%%%%%%%%%%%%%%%%%%%%%%%%%%%%%%%
We define the pseudoscalar decay constant $f_{PS}$ through the matrix element
\bee
\langle 0| \bar u \gamma_0 \gamma_5 d |PS\rangle = m_{PS} f_{PS}
\label{eq:fpi}
\ee
(so $f_{PS} \simeq 132$ MeV),
while the vector meson decay constant $f_V$ of state $V$ is defined as
\bee
\langle 0| \bar u \gamma_i d  | V\rangle = m_V^2 f_V \epsilon_i ,
\label{eq:fv}
\ee
where $\vec\epsilon$ is a polarization vector.
With clover fermions in the usual ($\kappa$) normalization, a continuum matrix element (carrying dimension $D$)
is defined to be
\bee
{\svev{\bar \psi \Gamma \psi}}_{\text{cont}} = \left(1 - \frac34\frac{\kappa}{\kappa_c}\right) Z_\Gamma
{\svev{\bar \psi \Gamma \psi}}_{\text{latt}} a^D,
\ee
and in perturbation theory for fermions in representation $R$
the one-loop renormalization factor is
\bee
Z_\Gamma = 1 + \frac{g^2 C_2(R)}{16\pi^2} z_\Gamma + \cdots.
\ee
$z_\Gamma$ for nHYP clover fermions is recorded in Ref.~\cite{DeGrand:2002vu}
 as $-1.28$ for the vector current and $-1.30$ for the axial current. In the usual tadpole-improved
analysis, one might take the coupling from the lowest-order expression for the plaquette,
using the fundamental representation Casimir,
\bee
-\Tr \frac{U_P}{N_c} = g^2 \frac{C_2(F)}{4}.
\ee
For the quenched data sets, the plaquette values (1.787, 2.858 and 3.976) give $g^2 C_2(F)$= 2--2.26.
In principle, we should run the scale of the coupling from its value for the plaquette, $q^*a=3.41$,
to the values computed in Ref.~\cite{DeGrand:2002vu}, $q^* a \simeq 1.7$, but the combination of coupling
and $z_\Gamma$ is so small for nHYP fermions that, in all cases, $Z_\Gamma$ is within 1\% of unity.

A first determination of $\kappa_c$ was described in Sec.~\ref{sec:phase}.
Figure \ref{fig:r0r1} shows that the lattice spacing is rather strongly dependent on
 $\kappa$ at fixed $\beta$,
so one would not expect a naive extrapolation of, say, $am_q$ or $(am_{PS})^2$ as a
 linear function of $\kappa$
would perform particularly well. In fact, it does not;
we can imagine doing fits to all four data points, or to the lightest three.
Since the fits have a nonzero number of degrees of freedom, we can evaluate their quality.
It is poor.

\begin{figure}[h]
\begin{center}
\includegraphics[width=\columnwidth,clip]{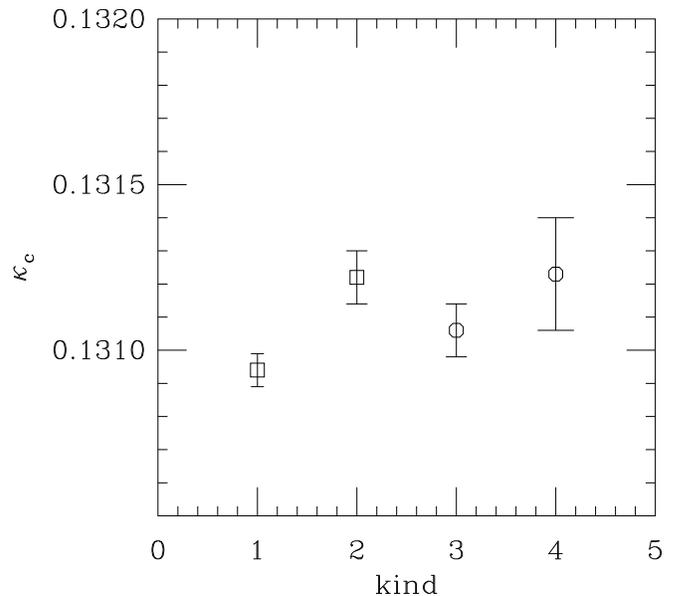}
\end{center}
\caption{Different determinations of $\kappa_c$ in the SU(4) AS2 theory: (1) from 
$r_1 m_q$ with all four mass values,
(2) from $r_1 m_q$ with the lowest three mass values,
(3) from $r_1^2 m_{PS}^2$ with all four mass values, and 
(4) from $r_1^2 m_{PS}^2$ with with the lowest three mass values.
\label{fig:kcerr}}
\end{figure}

Instead, we focus on the dimensionless quantities
 $r_1 m_q$ and $r_1^2 m_{PS}^2$.
A comparison of critical hopping parameters from the fits is shown in Fig.~\ref{fig:kcerr} 
for four possibilities,
all with a linear fit:
\begin{enumerate}
\item From $r_1 m_q$ with all four mass values [$\chi^2=24.5$ with 2 degrees of freedom (dof)];
\item From $r_1 m_q$ with the lowest three mass values ($\chi^2=2.4$ with 1 dof);
\item From $r_1^2 m_{PS}^2$ with all four mass values ($\chi^2=8.6$ with 2 dof);
\item From $r_1^2 m_{PS}^2$ with with the lowest three mass values ($\chi^2=7$ with 1 dof).
\end{enumerate}
The estimates of $\kappa_c$ are all quite close. More importantly, the uncertainty in
the rescaling between lattice and continuum-normalized matrix elements due to
different choices of $\kappa_c$ is under 0.5\% at any of the quark masses in our data sets.
The plots below assume $\kappa_c=0.13122$ and $Z=1$.

The partially quenched data sets, at fixed $\beta$ and sea quark $\kappa$,
should all have the same lattice spacing.
We should be able to find a ``valence $\kappa_c$'' just by fitting
 $am_q$ or $(am_{PS})^2$ to a straight line.
This we do, finding $\kappa_c=0.13137$.

The dynamical SU(3) data sets have $\kappa_c=0.12838(9)$ from a linear fit of
 $r_1 m_q$ in $\kappa$. The fit is stable with a $\chi^2$ below 1.1 per degree of freedom
 for the lowest five masses (or fewer).

%%%%%%%%%%%%%%%%%%%%%%%%%%%%%%%%%%%%%%%%%%%%%%%%%%%%%%%%%%%%%%%%%%%%%
%%%%%%%%%%%%%%%%%%%%%%%%%%%%%%%%%%%%%%%%%%%%%%%%%%%%%%%%%%%%%%%%%%%%%

\begin{figure}
\begin{center}
\includegraphics[width=\columnwidth,clip]{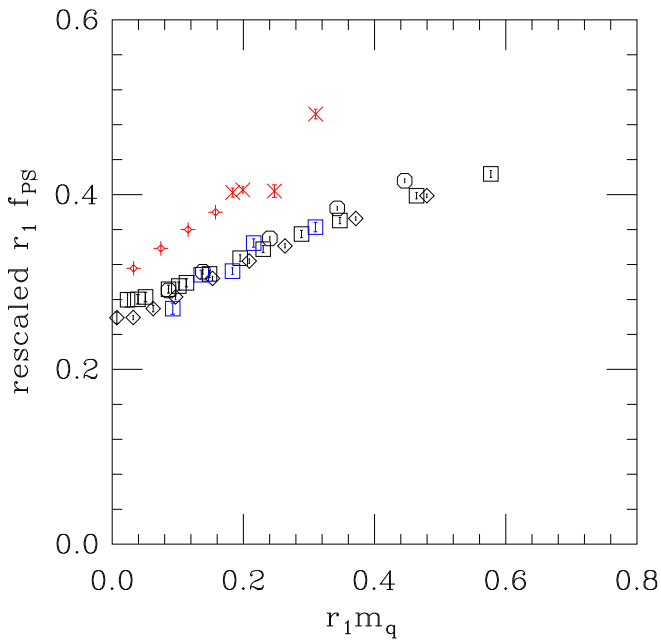}
\end{center}
\caption{Pseudoscalar decay constant.
 The abscissa is $r_1$ times the AWI quark mass.
The data sets are
black squares for quenched SU(3) fundamentals, black diamonds for quenched SU(5) fundamentals,
black octagons for quenched SU(7) fundamentals, and
red crosses for SU(4) AS2; the fancy diamonds are the PQ data. Finally, the blue squares
are  SU(3) with two dynamical, fundamental flavors.
\label{fig:fpi}}
\end{figure}

\begin{figure}[t]
\begin{center}
\includegraphics[width=\columnwidth,clip]{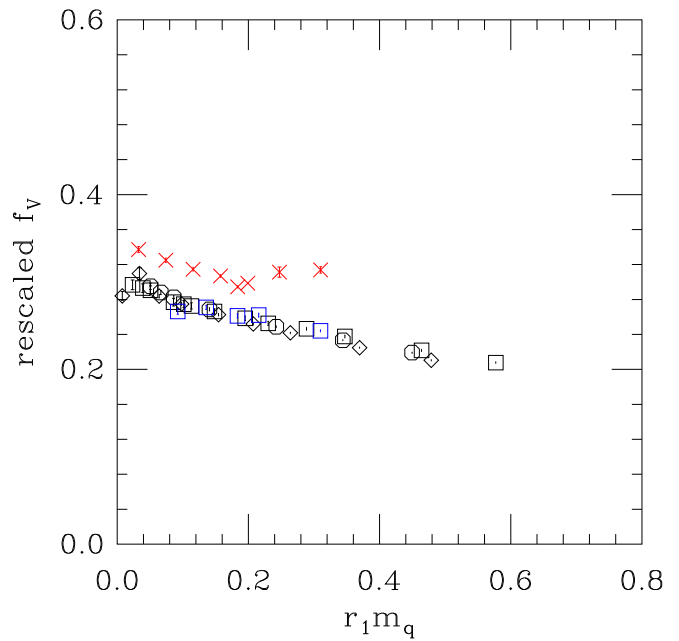}
\end{center}
\caption{Vector meson decay constant
 The abscissa is $r_1$ times the AWI quark mass.
The data sets are labeled as in Fig.~{\protect{\ref{fig:fpi}}}.
\label{fig:fv}}
\end{figure}

We collect our results for $f_{PS}$ and $f_V$ in Figs.~\ref{fig:fpi} and~\ref{fig:fv}.
We rescaled all fundamental-representation data by $\sqrt{3/N_c}$, and we rescaled the AS2 data by
$(3/N_c)$, to remove the leading expected large-$N_c$ scaling \cite{Cherman:2009fh}, leaving the residual. The
dynamical SU(3) data sets agree with the previously presented quenched sets (at the relatively heavy
quark masses where they overlap), and the trend of remarkable
$N_c$ scaling for the fundamental-representation data contrasts with the AS2 data sets,
where the shift from $N_c=3$ to 4 is about 20\%.  

 The slope of the rescaled $r_1 f_{PS}$ with respect
 to $r_1 m_q$ is roughly 50\% larger for the SU$(4)$ AS2 results, compared to all other results shown.  Next-to-leading-order
 chiral perturbation theory predicts a larger contribution by a factor of 2 from the
 low-energy constant $L_4$ for the SU$(4)$ AS2 data \cite{Bijnens:2009qm}; however,
 $L_4$ itself is usually taken to be small or even zero in QCD, since it is suppressed 
at large $N_c$ (with fundamental fermions) by the Okubo-Zweig-Iizuka (OZI)
 rule \cite{Gasser:1984gg}.\footnote{Recent global analyses of the low-energy constants
 in QCD \cite{Bijnens:2014lea} indicate that $L_4$ is not necessarily small, despite the
 expected large-$N_c$ suppression.}  The OZI rule, which follows from suppression
 of quark loops, does not hold for the AS2 expansion \cite{Cherman:2006iy} and so we might
 expect a larger slope for $f_{PS}$ vs $m_q$ in any AS2 theory compared to the conventional
 expansion at large $N_c$.  Results at larger values of $N_c$ with AS2 fermions would shed 
light on this discrepancy.

%%%%%%%%%%%%%%%%%%%%%%%%%%%%%%%%%%%%%%%%%%%%%%%%%%%%%%%%%%%%%%%%%%%%%
%%%%%%%%%%%%%%%%%%%%%%%%%%%%%%%%%%%%%%%%%%%%%%%%%%%%%%%%%%%%%%%%%%%%%

%%%%%%%%%%%%%%%%%%%%%%%%%%%%%%%%%%%%%%%%%%%%%%%%%%%%%%%%%%%%%%%%%%%%%
%%%%%%%%%%%%%%%%%%%%%%%%%%%%%%%%%%%%%%%%%%%%%%%%%%%%%%%%%%%%%%%%%%%%%

\pagebreak

%%%%%%%%%%%%%%%%%%%%%%%%%%%%%%%%%%%%%%%%%%%%%%%%%%%%%%%%%%%%%%%%%%%%%
\section{Comparisons: Baryons \label{sec:baryons}}
\begin{figure}
\begin{center}
\includegraphics[width=\columnwidth,clip]{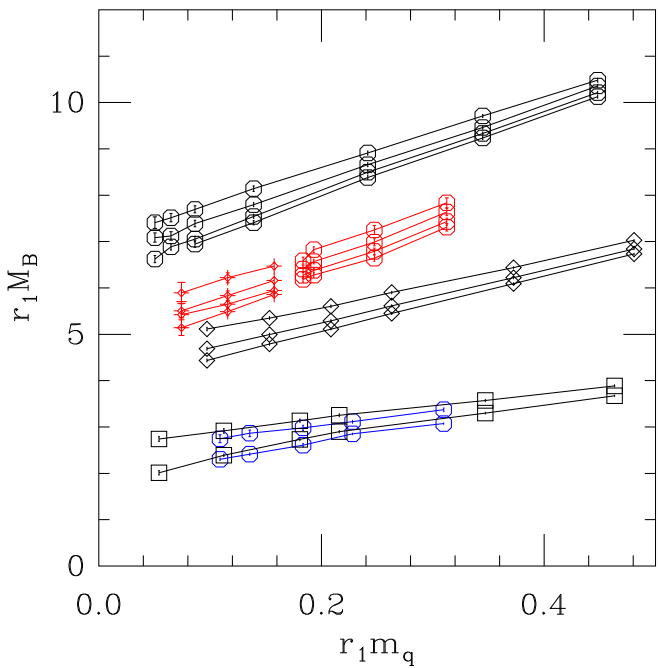}
\end{center}
\caption{Baryons. The black data are from the top quenched SU(7), SU(5) and SU(3) data.
The blue octagons are SU(3) with dynamical fermions. The red points are the six-quark baryons in SU(4) AS2,
octagons for dynamical data sets and fancy diamonds for the partially-quenched data sets.
\label{fig:baryons}}
\end{figure}

%%%%%%%%%%%%%%%%%%%%%%%%%%%%%%%%%%%%%%%%%%%%%%%%%%%%%%%%%%%%%%%%%%%%%
Our baryon data are shown in Fig.~\ref{fig:baryons}.
%%%%%%%%%%%%%%%%%%%%%%%%%%%%%%%%%%%%%%%%%%%%%%%%%%%%%%%%%%%%%%%%%%%%%
%%%%%%%%%%%%%%%%%%%%%%%%%%%%%%%%%%%%%%%%%%%%%%%%%%%%%%%%%%%%%%%%%%%%%
Unlike mesons, baryon masses depend strongly on $N_c$ and representation.
 Fundamental
representation data with
$N_c=3$, 5, and~7 make that point. In the figure, quenched data are shown in black while the blue points are
the SU(3) dynamical-fermion data.
 Again, we scale the lattice masses by $r_1$ and plot
the data versus the AWI quark mass. The SU(4) AS2 masses are shown in red,
with octagons for dynamical data sets and fancy diamonds for the partially quenched ones.
We have used the same symbols for all states, regardless of their angular momentum, but have
connected the states with the same $J$ by lines.
 The masses of all the states (all $N_c$, all representations) are ordered in angular
momentum so that higher $J$ lies higher.
Of course, the masses in each set come from the same underlying configurations, so they are highly correlated
and move together as the quark mass is varied.

\begin{figure}
\begin{center}
\includegraphics[width=\columnwidth,clip]{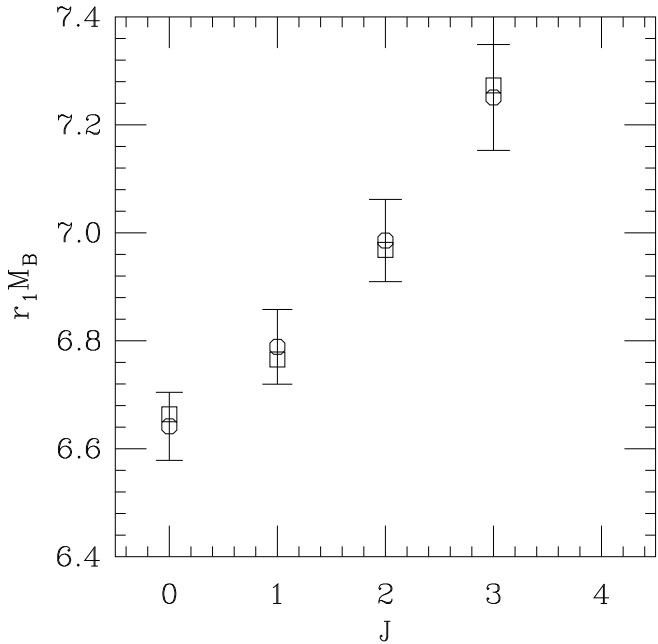}
\end{center}
\caption{Fit to rotor formula (\protect{\ref{eq:jsplit}}) at $\kappa=0.1285$.
Octagons (with error bars) are the data points;  squares
the best fit values.
\label{fig:f01285}}
\end{figure}

\begin{figure*}
\begin{center}
\includegraphics[width=\textwidth,clip]{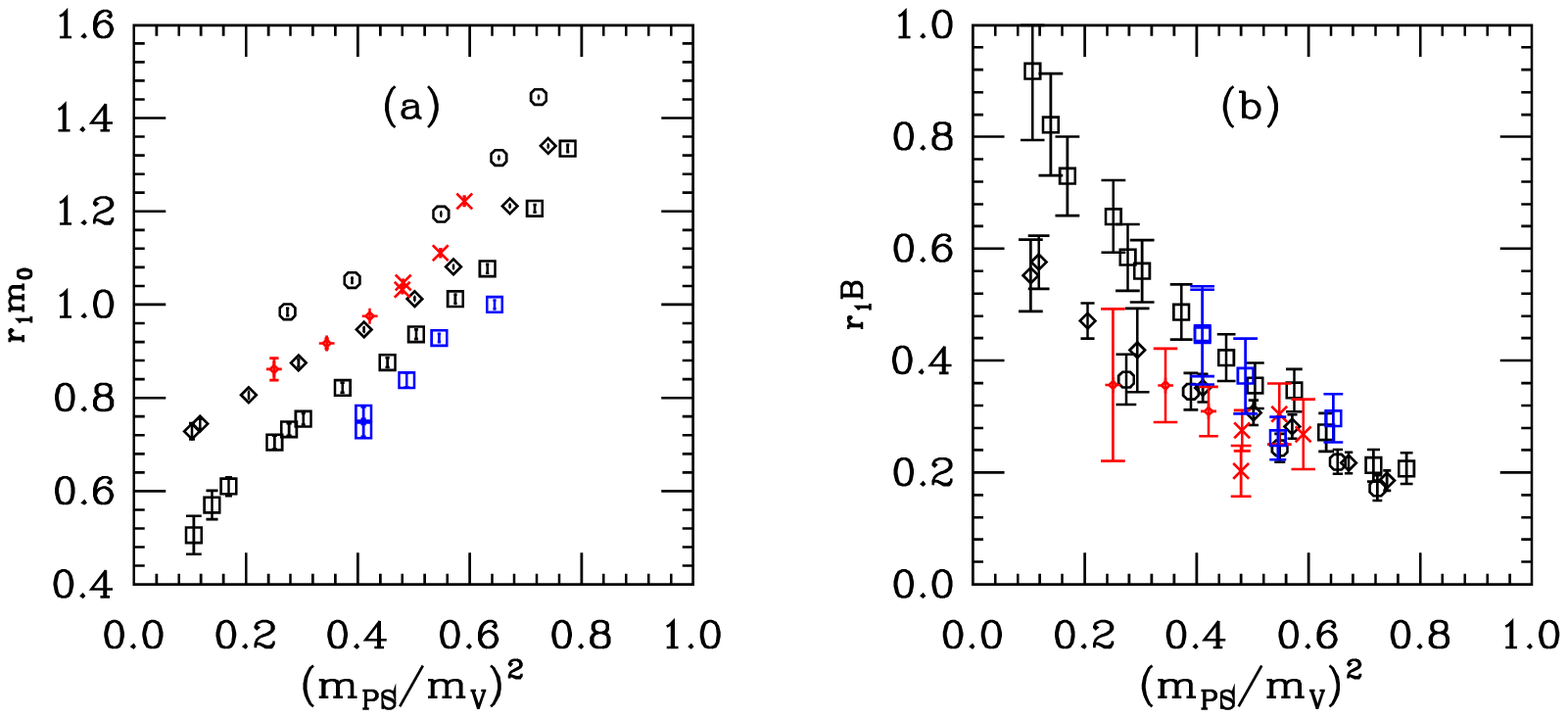}
\end{center}
\caption{
The parameter $m_0$ [panel (a)] and $B$ [panel (b)]
from  two-flavor degenerate mass data as  function of $(m_{PS}/m_V)^2$ from a fit to
\Eq{eq:jsplit}.
Data from the quenched SU(3), SU(5), and SU(7) multiplets are
shown respectively
as squares, diamonds, and octagons. Red crosses and fancy diamonds show the SU(4) data,
unquenched and partially quenched, and
the blue squares are the dynamical SU(3) data sets.
\label{fig:pvspirho}}
\end{figure*}

We can compare the fine structure in the AS2 data to the familiar rotor formula~\cite{Adkins:1983ya,Jenkins:1993zu,Buisseret:2011aa},
\bee
M_B(J)= m_0 N_b + B\frac{J(J+1)}{N_b}.
\label{eq:jsplit}
\ee
$N_b$ is the number of quarks in the baryon, and $m_0$ can be interpreted as a constituent quark mass.
Thus we set $N_b=N_c$ for fundamental-representation fermions and  $N_b=6$ for SU(4) AS2\@.
Equation~(\ref{eq:jsplit}) describes the data well.
This is shown  for one quark mass, $\kappa=0.1285$, in Fig.~\ref{fig:f01285}.
%%%%%%%%%%%%%%%%%%%%%%%%%%%%%%%%%%%%%%%%%%%%%%%%%%%%%%%%%%%%%%%%%%%%%
%%%%%%%%%%%%%%%%%%%%%%%%%%%%%%%%%%%%%%%%%%%%%%%%%%%%%%%%%%%%%%%%%%%%%
The masses of the four different $J$ states are fit to two parameters, $m_0$ and $B$.
The results of the fit are shown as squares in the figure.
Repeating these fits for all masses, we can plot the quark mass dependence of $m_0$ and $B$.
This is shown in Fig.~\ref{fig:pvspirho}.
%%%%%%%%%%%%%%%%%%%%%%%%%%%%%%%%%%%%%%%%%%%%%%%%%%%%%%%%%%%%%%%%%%%%%
%%%%%%%%%%%%%%%%%%%%%%%%%%%%%%%%%%%%%%%%%%%%%%%%%%%%%%%%%%%%%%%%%%%%%

Some residual $N_c$ dependence is observed in Fig.~\ref{fig:pvspirho}, especially in $m_0$,
shown in the left-hand panel.
This situation for  the quenched fundamental data was discussed in
Ref.~\cite{DeGrand:2013nna}. It was observed that the variation in the data was (noisily) consistent with
a $1/N_c$ contribution to $m_0$; that is, $m_0(N_c) = m_{00} + m_{01}/N_c + \dots$ where
$m_{00}$ and $m_{01}$ were of comparable, ``typical QCD'' size.  With only two AS2 points to compare, we cannot reliably fit for the corresponding $N_c$ dependence.  However, we observe that modeling $m_0(N_b) = m_{00} + m_{01}/N_b + \dots$ gives roughly consistent results with our $N_c=4$ AS2 data.

The situation for $B$ is less clear cut: $B$ comes from small mass differences. Certainly, the
$N_c=5$ and 7 fundamental $B$ data and the $N_c=4$ AS2 $B$ data lie on a common line slightly separated from the $N_c=3$ data. This is in qualitative agreement with large-$N_c$ expectations, $B(N_b) = B_0 + B_1/N_b + \cdots$.

\begin{figure}[t]
\begin{center}
\includegraphics[width=\columnwidth,clip]{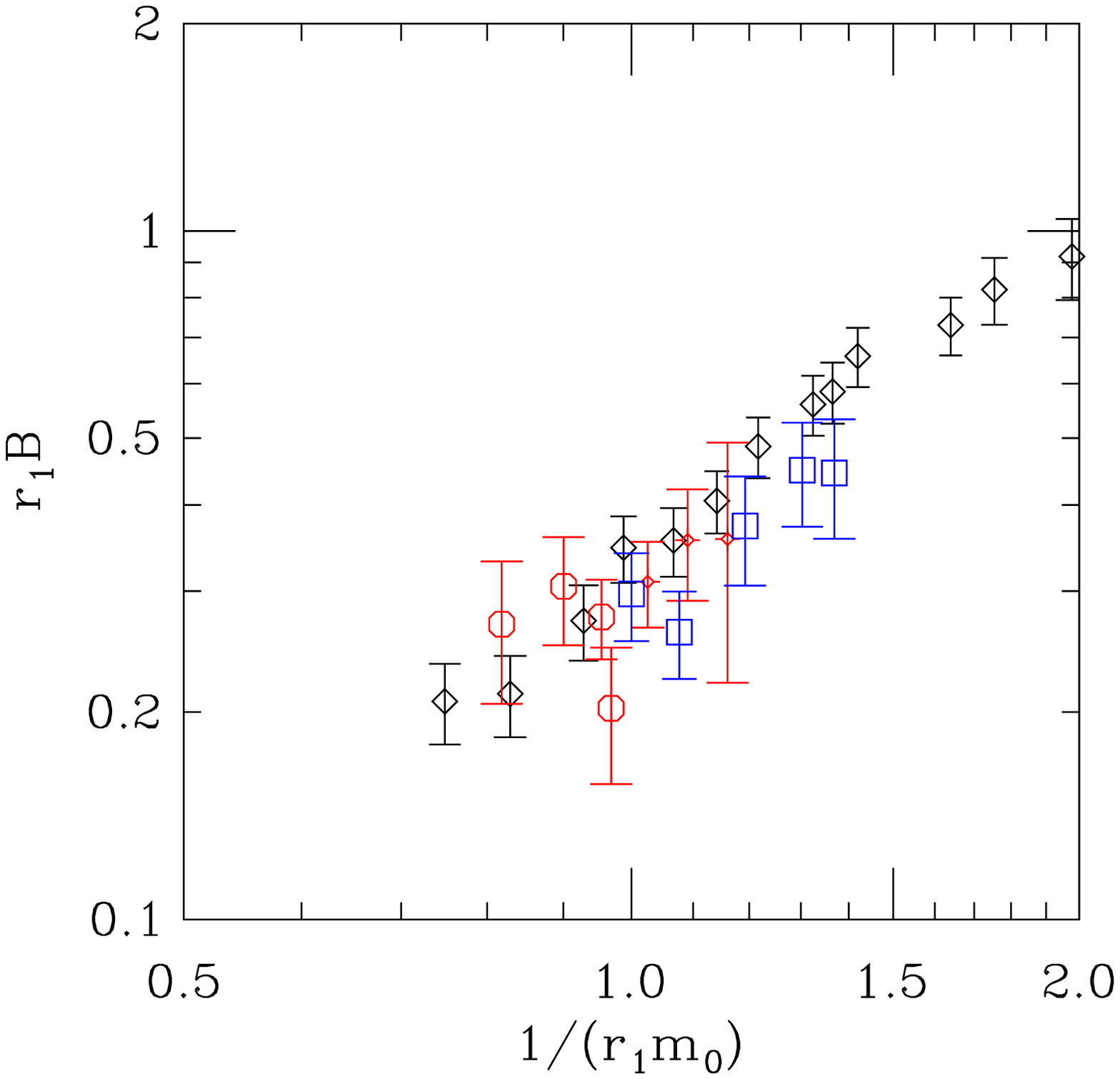}
\end{center}
\caption{$B$ vs $1/m_0$ from the rotor formula (\protect{\ref{eq:jsplit}});
 black diamonds from quenched SU(3), blue squares from full SU(3).
The SU(4) data are shown as red
octagons for the dynamical sets and fancy diamonds for the partially quenched set.
\label{fig:bvs1m0}}
\end{figure}

\begin{figure}[b]
\begin{center}
\includegraphics[width=\columnwidth,clip]{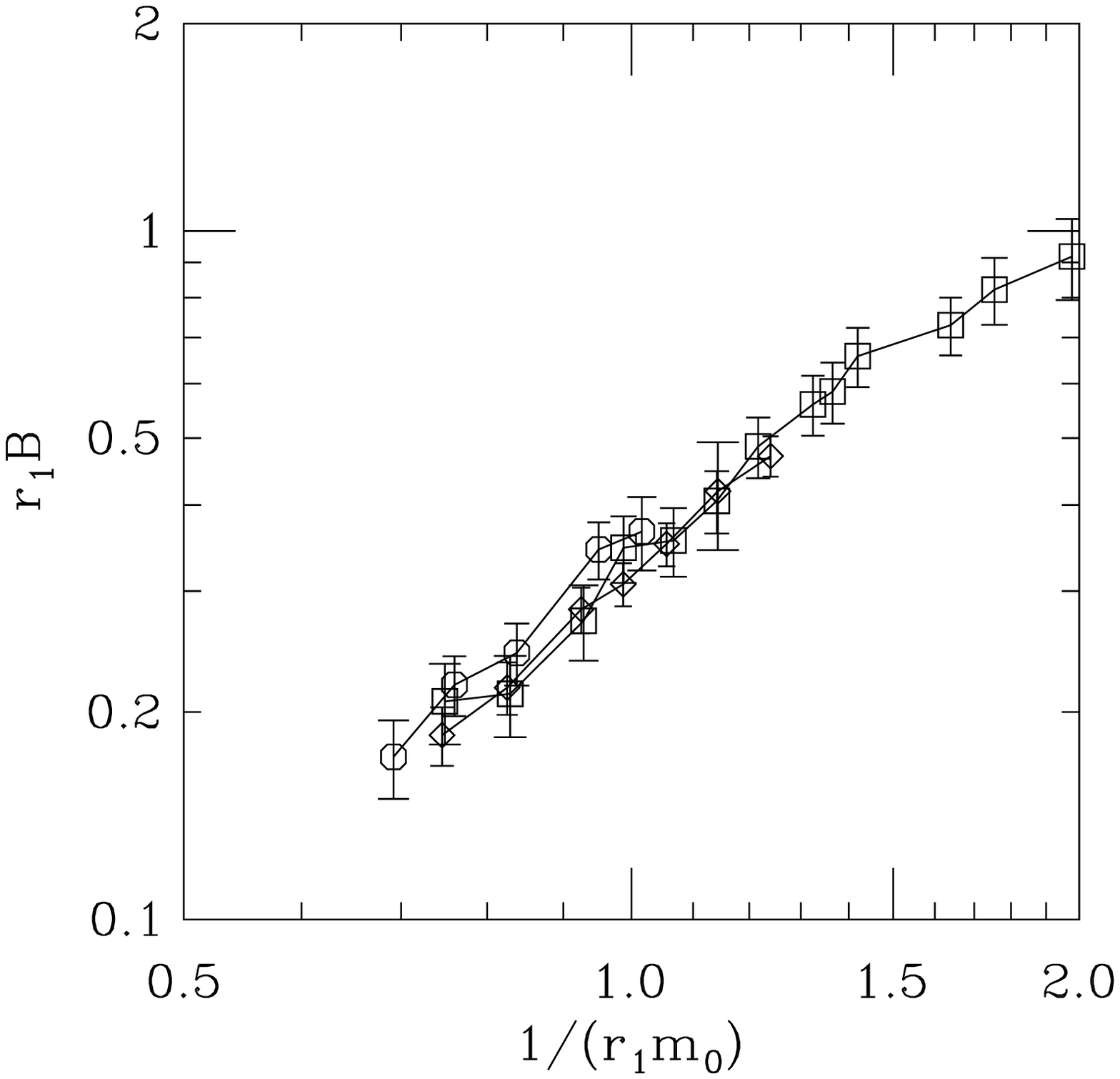}
\end{center}
\caption{For comparison, $B$ vs $1/m_0$ from the rotor formula (\protect{\ref{eq:jsplit}})
from the quenched fundamental data sets of Ref.~{\protect{\cite{DeGrand:2012hd}}}:
$N_c=3$, 5, and 7 data sets are squares, diamonds and octagons.
\label{fig:bvs1m0q}}
\end{figure}

Overall, both $m_0$ and $B$ are of ``typical hadronic size''
since $1/r_1\sim 635$ MeV and $r_1 m_0$ and $r_1 B$ are order unity. However, they have
rather different dependence on the quark mass. In the Skyrme picture, $B$ is the inverse
 of the moment of inertia,
scaled by $N_b$, so that $B$ should be proportional to $1/m_0$. In a quark model with hyperfine interactions
mediated by gluons, $B$ is basically a product of color magnetic moments for the quarks,
and for heavy quarks, the magnetic moment scales inversely with the quark mass. This
suggests $B\propto 1/m_0^2$. A log-log plot of $B$ versus $1/m_0$ certainly looks like a power law,
 with an exponent near unity. This is shown for $N_c=3$ and~4 in Fig.~\ref{fig:bvs1m0}
and for the quenched fundamental data in Fig.~\ref{fig:bvs1m0q}.

%%%%%%%%%%%%%%%%%%%%%%%%%%%%%%%%%%%%%%%%%%%%%%%%%%%%%%%%%%%%%%%%%%%%%
%%%%%%%%%%%%%%%%%%%%%%%%%%%%%%%%%%%%%%%%%%%%%%%%%%%%%%%%%%%%%%%%%%%%%
%%%%%%%%%%%%%%%%%%%%%%%%%%%%%%%%%%%%%%%%%%%%%%%%%%%%%%%%%%%%%%%%%%%%%
%%%%%%%%%%%%%%%%%%%%%%%%%%%%%%%%%%%%%%%%%%%%%%%%%%%%%%%%%%%%%%%%%%%%%

The overall dependence of the baryon mass $M_B$ on the quark mass $m_q$ is also 
interesting to study, since it may be used with the Feynman-Hellmann theorem to 
determine the baryonic matrix element of the scalar density, if one defines
\begin{equation}\label{eq:barysigma}
f_q^{(B)} \equiv \frac{m_q}{M_B} \frac{\partial M_B}{\partial m_q} = 
\frac{m_q}{M_B} \langle B | \bar{\psi} \psi | B \rangle.
\end{equation}
Multiplying by the ratio $m_q/M_B$ cancels the renormalization of the quark mass and 
gives a dimensionless ratio.
 For the lowest-lying baryon, this quantity would determine the cross 
section for direct detection through Higgs exchange in the context of a composite dark matter
 model \cite{Appelquist:2014jch}, in conjunction with the same quantity defined for matrix
 elements of the proton and neutron \cite{Junnarkar:2013ac}.

To determine the scalar matrix element, we carry out a linear fit to the quantity $r_1 M_B$ as
 a function of $r_1 m_q$; the resulting slope is then multiplied by $m_q / M_B$ at each data
 point.  To suppress possible finite-volume systematic errors, only points
 with $m_{PS} N_s \gtrsim 4$ are used in the fit; this excludes a small fraction of our data.

\begin{figure}[b]
\begin{center}
\includegraphics[width=\columnwidth,clip]{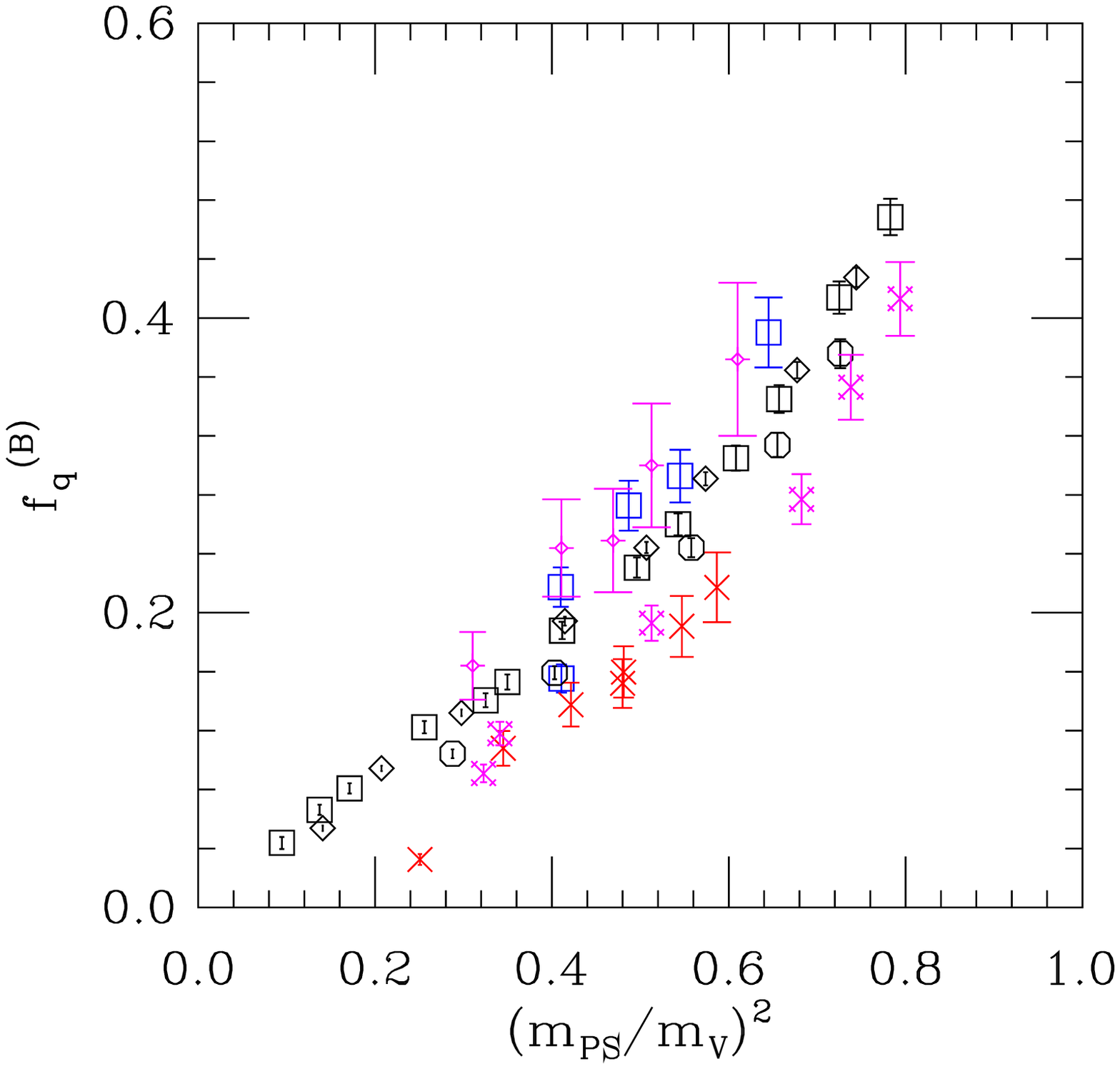}
\end{center}
\caption{
The quantity $f_q^{(B)}$ defined in Eq.~(\ref{eq:barysigma}), plotted vs the ratio $(m_{PS}/m_V)^2$. 
 Data shown include quenched fundamental SU$(3)$, SU$(5)$, and SU$(7)$ (black squares, diamonds, and octagons),
 dynamical SU$(3)$ (blue squares), and dynamical SU$(4)$ AS2 (red crosses).  We also plot in purple
 results from Ref.~\cite{Appelquist:2014jch} for quenched fundamental SU$(4)$, for bare gauge 
coupling $\beta = 11.5$ (fancy diamonds) and $\beta = 12.0$ (fancy crosses).
\label{fig:barysigma}}
\end{figure}

Results of this analysis are plotted in Fig.~\ref{fig:barysigma}.  Since in the limit
 $m_q \rightarrow \infty$ we expect $M_B \sim N_b m_q$, the quantity $f_q^{(B)}$ should
 approach 1 in the heavy-quark limit and 0 in the chiral limit.  The functional dependence observed for all of our AS2 and fundamental data is broadly consistent at intermediate values of $m_{PS}/m_V$, and consistency is also seen with other lattice results for SU(2) and SU(4) theories with
 relatively heavy quark masses \cite{Appelquist:2014jch,Detmold:2014kba}.

%%%%%%%%%%%%%%%%%%%%%%%%%%%%%%%%%%%%%%%%%%%%%%%%%%%%%%%%%%%%%%%%%%%%%
%%%%%%%%%%%%%%%%%%%%%%%%%%%%%%%%%%%%%%%%%%%%%%%%%%%%%%%%%%%%%%%%%%%%%

%%%%%%%%%%%%%%%%%%%%%%%%%%%%%%%%%%%%%%%%%%%%%%%%%%%%%%%%%%%%%%%%%%%%%
\section{Consequences and Conclusions\label{sec:conclusions}}
%%%%%%%%%%%%%%%%%%%%%%%%%%%%%%%%%%%%%%%%%%%%%%%%%%%%%%%%%%%%%%%%%%%%%
We have presented a first lattice calculation of the spectrum for an SU(4) gauge theory with two Dirac fermions in the two-index antisymmetric (AS2) representation.  Because this is a real representation, its symmetries are somewhat different from the familiar QCD case; in particular the chiral symmetry group is enlarged, breaking SU(4) $\rightarrow$ SO(4).  We have clarified some features of this symmetry, particularly as relevant for lattice simulations.  Furthermore, we have mapped out the phase diagram for our lattice action, and identified and removed a novel discretization error in nHYP smearing which appears for real-representation fermions.  Our work provides a foundation for future studies of SU(4) theories with AS2 fermions, and for other lattice studies of theories with real-representation fermions.

Comparisons of SU(4) AS2 spectroscopy and matrix elements with fundamental fermion data
reveal regularities anticipated by large-$N_c$ arguments.  Although we cannot derive
 any quantitative results on the nature of the AS2 large-$N_c$ expansion with 
only two points ($N_c = 3$ and $N_c = 4$), our results for meson and baryon masses seem
consistent with the predictions of the large-$N_c$ framework.

Scaling of decay constants seems to be less exact for AS2 fermions than for fundamental ones, with the SU(4) AS2 results clearly distinct from the various SU($N_c$) fundamental theories.  We note, however, that for $N_c\ne 4$, AS2 fermions live in complex representations. The pattern of chiral symmetry breaking is then identical to that of ordinary QCD.  Since $N_c= 4$ is a special case, it might be an outlier for the behavior of chirally sensitive observables such as $f_{PS}$.  

As far as we know, no dynamical
simulations of gauge plus fermionic systems on volumes large enough for spectroscopy
with $N_c>4$ have ever been performed.
At heavier quark masses, however, quenching effects are not large.
For future study, perhaps it would be appropriate to imagine a first round of quenched simulations with AS2 fermions.
 Studies of mesonic properties could be done with modest resources.
With $N_b=N_c(N_c-1)/2$ quarks in a baryon,
they are bosons for $N_c=5$  (with 10 constituents), and they alternate between fermion and
boson at larger
$N_c$. 
Unfortunately, the number of terms in the wave function grows rapidly with $N_c$.
Even for $N_c=5$, the combinatorics of the lower-$J$ correlators seem quite daunting.  

Returning to the $N_c=4$ theory, the six-quark AS2 baryons are almost certainly unstable against decay in the chiral limit, since they can fall apart into three diquarks.  As an example, consider any of our baryons with $I=J>0$. Diquark NGBs have $I=1$, and so the decay into three such NGBs is allowed by isospin conservation.  However, the NGBs have $J=0$, meaning
that the baryon's angular momentum will have to be converted into an orbital
motion.  This leads to a kinematic suppression of the decay.
The same applies to the $I=J=0$ baryon:  The isospin state of the three NGBs is antisymmetric, 
so they will have to be in a spatially antisymmetric state that perforce contains orbital angular momentum.
Of course, knowing that the decay actually occurs
as a strong-interaction process might be sufficient for
phenomenology.  %For example, in the chiral limit of a model like this, 
%the lightest baryon would probably not be a good dark matter candidate because
%it would decay into massless NGBs. 

As mentioned in the Introduction, of particular interest for phenomenology
is the $N_{\text{Maj}}=5$ theory, the low-energy effective theory of which
is the SU(5)/SO(5) nonlinear sigma model.  We have begun a detailed study
of this model that we hope to report on in the future.

\pagebreak

%%%%%%%%%%%%%%%%%%%%%%%%%%%%%%%%%%%%%%%%%%%%%%%%%%%%%%%%%%%%%%%%%%%%%
\begin{acknowledgments}
T.~D. would like to thank Richard Lebed for correspondence and conversations.
B.~S. thanks the University of Colorado for hospitality, as well as the Yukawa
 Institute for Theoretical Physics at Kyoto University.
This work was supported in part by the U.~S. Department of Energy, and by the Israel Science 
Foundation under Grant no.~449/13.  Brookhaven National Laboratory is supported by the U.~S.~Department of Energy under contract DE-SC0012704.
Computations were performed using USQCD resources at Fermilab and
on the University of Colorado theory group's cluster.
Our computer code is based on version 7 of the publicly available code of the MILC
collaboration~\cite{MILC}.
\end{acknowledgments}

\clearpage

%%%%%%%%%%%%%%%%%%%%%%%%%%%%%%%%%%%%%%%%%%%%%%%%%%%%%%%%%%%%%%%%%%%%%

%%%%%%%%%%%%%%%%%%%%%%%%%%%%%%%%%%%%%%%%%%%%%%%%%%%%%%%%%%%%%%%%%%%%%
\end{document}